\documentclass[twocolumn]{aastex6}
\pdfoutput=1

\usepackage{amssymb}
\usepackage{amsmath}            		                 		
\usepackage{graphicx}
\usepackage{textcomp}
\usepackage{wrapfig}
\usepackage{url}
\usepackage{float}
\usepackage[utf8]{inputenc}
\usepackage[T1]{fontenc}

\def\NII  {[N{\small{II}}]}
\def\CII  {[C{\small{II}}]}
\def\HII  {H{\small{II}}}
\def\OI  {[O{\small{I}}]}

\def\micron {$\mu$m}

\begin{document}

\title{Using \CII~158~\micron\ Emission From Isolated ISM Phases as a Star--Formation Rate Indicator}
\author{Jessica Sutter\altaffilmark{1},
Daniel A. Dale\altaffilmark{1}, 
Kevin V. Croxall \altaffilmark{2},
Eric W. Pelligrini \altaffilmark{3},
J.D.T. Smith\altaffilmark{4},
Philip N. Appleton \altaffilmark{5}
Pedro Beir{\~a}o \altaffilmark{6},
Alberto D. Bolatto\altaffilmark{7},
Daniela Calzetti\altaffilmark{8},
Alison Crocker \altaffilmark{9},
Ilse De Looze\altaffilmark{10},
Bruce Draine \altaffilmark{11},
Maud Galametz\altaffilmark{12},
Brent A. Groves\altaffilmark{13},
George Helou\altaffilmark{5},
Rodrigo Herrera-Camus\altaffilmark{14},
Leslie K. Hunt\altaffilmark{15},
Robert C. Kennicutt\altaffilmark{16, 17},
H\'el\`ene Roussel \altaffilmark{18},
Mark G. Wolfire\altaffilmark{4}
}
\altaffiltext{1}{Department of Physics \& Astronomy, University of Wyoming, Laramie WY; jsutter4@uwyo.edu}
\altaffiltext{2}{Expeed Software, Columbus, OH}
\altaffiltext{3}{Institute for Theoretical Astrophysics Heidelberg, Germany}
\altaffiltext{4}{Department of Physics \& Astronomy, University of Toledo, Toledo, OH}
\altaffiltext{5}{IPAC, California Institute of Technology, Pasadena, CA}
\altaffiltext{6}{Machine Learning Company, Oss, The Netherlands}
\altaffiltext{7}{Department of Astronomy, University of Maryland, College Park, MD}
\altaffiltext{8}{Department of Astronomy, University of Massachusetts, Amherst MA}
\altaffiltext{9}{Physics Department, Reed College, Portland OR}
\altaffiltext{10}{Department of Physics \& Astronomy, University College London, London, UK}
\altaffiltext{11}{Department of Astrophysical Sciences, Princeton University, Princeton, NJ}
\altaffiltext{12}{Laboratoire AIM, CEA, Universit\'{e} Paris Diderot, IRFU/Service d'Astrophysique, Gif-sur-Yvette, France}
\altaffiltext{13}{Research School of Astronomy \& Astrophysics, Australian National University, Canberra, Australia}
\altaffiltext{14}{Departmento de Astronom\'ia, Facultad de Ciencias F\'isicas y Matem\'aticas, Universidad de Concepci\'on, Concepci\'on, Chile}
\altaffiltext{15}{INAF -  Osservatorio Astrofisico di Arcetri, Firenze, Italy}
\altaffiltext{16}{Steward Observatory, University of Arizona, Tucson, AZ}
\altaffiltext{17}{Department of Physics \& Astronomy, Texas A\&M University, College Station, TX}
\altaffiltext{18}{Institut d'Astrophysique de Paris, Sorbonne Universit\'es, Paris, France}

\begin{abstract}
The brightest observed emission line in many star-forming galaxies is the \CII~158~\micron\ line, making it detectable up to z$\sim$7.  In order to better understand and quantify the \CII\ emission as a tracer of star-formation, the theoretical ratio between the \NII~205~\micron\ emission and the \CII~158~\micron\ emission has been employed to empirically determine the fraction of \CII\ emission that originates from the ionized and neutral phases of the ISM.  Sub-kiloparsec measurements of the  \CII~158~\micron\ and \NII~205~\micron\ line in nearby galaxies have recently become available as part of the Key Insights in Nearby Galaxies: a Far Infrared Survey with \textit{Herschel} (KINGFISH) and Beyond the Peak (BtP) programs.  With the information from these two far-infrared lines along with the multi-wavelength suite of KINGFISH data, a calibration of the \CII\ emission line as a star formation rate indicator and a better understanding of the \CII\ deficit are pursued.  \CII\ emission is also compared to PAH emission in these regions to compare photoelectric heating from PAH molecules to cooling by \CII\ in the neutral and ionized phases of the ISM.  We find that the \CII\ emission originating in the neutral phase of the ISM does not exhibit a deficit with respect to the infrared luminosity and is therefore preferred over the \CII\ emission originating in the ionized phase of the ISM as a star formation rate indicator for the normal star-forming galaxies included in this sample. 
\end{abstract}

\section{Introduction} \label{sec:intro}

The \CII~158~\micron\ line is frequently the brightest \textit{observed} emission line in star-forming galaxies \citep{Luhman2003, Brauher2008}.  This prominence is due to the prevalence of carbon, its dominant role in cooling neutral atomic gas \citep{Wolfire2003}, its sub-Rydberg ionization potential, and the minimal dust attenuation it undergoes given its long wavelength \citep{Luhman1998, Heiles1994}.  The brightness of this line makes \CII~158~\micron\ emission an invaluable tool for probing the interstellar medium (ISM) of remote galaxies and the fainter regions within the disks of nearby galaxies.  As \CII~158~\micron\ is a prominent and typically unattenuated emission line, it has naturally become a target of interest for tracing physical properties such as star formation rate (SFR).  \CII\ emission is expected to trace star formation as \CII\ is a primary coolant of the photodissociation regions (PDRs), meaning cooling by \CII\ emission should balance photoelectric heating from young, hot stars to maintain thermal stability.   However, the relatively low ionization potential of neutral carbon (11.3~eV) complicates the potential diagnostic capabilities of this line.  Due to its low ionization potential, C$^+$ can exist within both the ionized and neutral ISM, including ISM phases spanning \HII\ regions, warm ionized gas, cold atomic gas, and PDRs, which effects the interpretive power of the 158~\micron\ emission line as a tracer of any specific galactic property \citep{Stacey1985, Shibai1991, Bennett1994, Pineda2013}.  

This multiphase origin also affects \CII's potential as a SFR indicator through the effect known as the ``\CII\ deficit'' \citep{Malhotra2001, Croxall2018, HerreraCamus2015}.  The \CII\ deficit is the decreasing trend in \CII~158~\micron\ luminosity to total infrared (TIR) luminosity (i.e. the total luminosity over 5--1100~\micron) with respect to various measures of luminosity or star forming activity.  Multiple studies of  \CII~/~TIR have found a decrease in this ratio as a function of increasing star formation rate surface density \citep{Smith2017}, infrared luminosity \citep{Malhotra2001, Luhman2003}, far infrared color \citep{Helou2001, Croxall2012, DiazSantos2017}, and the ratio of infrared luminosity to H$_2$ gas mass ($L_{\rm IR} / M_{\rm H_{2}}$) \citep{GraciaCarpio2011, HerreraCamus2018a}.  Several mechanisms have been proposed and investigated to explain the observed relative decrease in \CII\ emission.  These include: (1) \CII\ is optically thick or absorbed by dust in galaxies with the highest infrared luminosities \citep{Abel2007, Neri2014}, (2) that AGN activity in the host galaxies could produce higher infrared luminosity and lower \CII\ emission due to the increased hardness of radiation in AGN host galaxies potentially changing the ionization states of carbon in the ISM \citep{Langer2015, HerreraCamus2018b}, (3) the thermalization and saturation of the \CII\ line in warm, high density environments leads to the \OI~63~\micron\ line becoming the dominant cooling line in these regions \citep{Munoz2016, DiazSantos2017}, (4) in regions with high ionization parameters, a majority of the FUV radiation from young stars is absorbed within the \HII\ regions rather than escaping to PDRs where \CII\ is the primary coolant \citep{GraciaCarpio2011, Abel2009}, or (5) that very small dust grains in the most infrared luminous galaxies become highly charged by increased FUV emission from star formation, increasing the energy needed to photo-eject additional electrons, reducing the number and energy of photo-ejected electrons per unit FUV radiation \citep{Malhotra2001, GraciaCarpio2011}.  The \CII\ deficit has been measured across a wide range of galaxy samples, from low-metallicity dwarf galaxies \citep{Cormier2019}, to Ultra/Luminous Infrared Galaxies with infrared luminosities above $10^{11} L_{\odot}$ (U/LIRGS) \citep{DiazSantos2017}, to normal star-forming galaxies \citep{Malhotra2001, Smith2017}. These studies of multiple fine-structure cooling lines in a variety of galaxies and others like them have indicated that the third, fourth, and fifth explanation seem to be the most prominent causes of this deficit, although all five may play some part in creating the observed decline in the \CII\ to TIR luminosity ratio \citep{Luhman2003, Malhotra2001, Smith2017, Croxall2012, DiazSantos2017}.

Despite these difficulties, multiple groups have attempted to use \CII\ alone as a star formation rate indicator \citep[e.g.,][]{Stacey1991, Boselli2002, DeLooze2011, HerreraCamus2015}.  Although there are indications that \CII\ emission primarily originates in and around star forming regions \citep{Stacey1985, Mookerjea2011}, the relationships found between \CII\ luminosity and other tracers of star formation like extinction--corrected H$\alpha$ \citep{Boselli2002} and FUV luminosity \citep{DeLooze2011} often show a large scatter (with scatter as large as a factor of ten) and do a poor job of matching star formation rates in extreme cases \citep{DiazSantos2017, DeLooze2014, Sargsyan2012}.  However, it has been found that by including additional spectral information it is possible to determine a better constrained \CII--SFR relationship.  For example, using \CII\ surface brightness along with an infrared color correction can predict the star-formation rate to within a factor of three for all but the most IR luminous systems \citep{HerreraCamus2015}.  Other work has shown that the ratio of \CII\ to TIR luminosity exhibits a strongly non-linear but quantifiable trend with SFR surface density \citep{Smith2017}.

In order to further decode any relationship between \CII~158~\micron\ emission and SFR, the \CII\ emission can be separated by ISM phase in which it originates.  The \NII~205~\micron\ line is a powerful tool for separating \CII\ emission into neutral and ionized ISM components.  As the \NII~205~\micron\ line is also a far infrared fine structure line, it is typically unattenuated by dust, similar to the \CII~158~\micron\ line.  Since neutral nitrogen has an ionization energy of 14.5~eV, slightly above the 13.6~eV ionization energy of hydrogen, N$^+$ mainly exists in environments where hydrogen is ionized; \NII~205~\micron\ emission is essentially from \HII\ regions and other ionized phases of the ISM only.  Due to this constraint on the origin of the \NII~205~\micron\ emission, \NII\ can be used to constrain the fraction of \CII\ emission arising from the ionized ISM.  Since the \NII~205~\micron\ line has a similar critical density as the \CII~158~\micron\ line \citep[$\sim$32~cm$^{-3}$ for \NII~205~\micron\ and $\sim$45~cm$^{-3}$ for \CII~158~\micron][]{Oberst2006, Croxall2017}, the ratio of \CII~158~\micron\ emission to \NII~205~\micron\ emission is nearly constant regardless of electron number density \citep[see Figure~\ref{fig:ciitonii}, and][] {Croxall2012}.  This consistency makes \NII~205~\micron--derived measures of the fraction of \CII~158~\micron\ emission from the ionized ISM less dependent on electron number density ($n_{e}$) than when using the brighter \NII~122~\micron\ line, as the ratio of the \CII~158~\micron\ and \NII~122~\micron\ lines varies by a factor of three in normal ISM density conditions \citep{Croxall2012}.  Therefore, \NII~205~\micron\ emission is the preferred tool to distinguish between the \CII\ originating from the ionized and neutral phases of the ISM.  

With the added information from the \NII~205~\micron\ line, the \CII~158~\micron\ emission can be separated by ISM phase.  Isolated ionized and neutral ISM phase \CII\ emission can then be calibrated and tested as indicators of SFR.  Unfortunately, due to the weakness of the \NII~205~\micron\ line and its location in the far-infrared part of the spectrum, it is notoriously difficult to detect in Local Universe galaxies.  Fortunately, a collection of \NII~205~\micron\ detections was made with the \textit{Herschel Space Observatory} \citep{Pilbratt2010, Poglitsch2010}.  

An intriguing application of an improved \CII--based SFR indicator lies in the realm of high redshift galaxies.  There have been many detections of the \CII~158~\micron\ line in galaxies from $z\sim1$ to $z\sim7$ \citep{HaileyDunsheath2010, Ivison2010, Stacey2010, GraciaCarpio2011, Valtchanov2011, Gullberg2015, Malhotra2017, Barisic2017, Gullberg2018, Rybak2019}.  The brightness of the \CII\ line enables detections across a wide range of distances, making it a popular tool for probing the PDR properties and kinematics of galaxies in the high-redshift Universe.  In addition to the availability of high redshift \CII\ detections, the Atacama Large Millimeter Array (ALMA) has already detected \NII\ in multiple galaxies beyond a redshift of four \citep{Decarli2014, Aravena2016, Pavesi2016, Lu2017}.  Other work has identified the \CII\ and \NII\ lines in larger samples of galaxies between $1<z<2$ \citep{Stacey2010}, where cosmic star formation peaks \citep{Madau2014}.  As the \CII~158~\micron\ emission line is now measurable both in the Local Universe and in distant galaxies, it can be used to trace star formation at nearly any cosmic epoch.  Also, unlike ultraviolet and optical star formation tracers such as FUV and H$\alpha$, attenuation by dust is typically negligible at 158~\micron.

This paper uses \CII~158~\micron\ and \NII~205~\micron\ measurements from the Key Insights on Nearby Galaxies: a Far-Infrared Survey with \textit{Herschel} (KINGFISH) \citep{Kennicutt2011} and Beyond the Peak (BtP) \citep{Pellegrini2013} data sets to decompose the \CII~158~\micron\ emission into the ISM phases in which it originates.  These samples include normal, Local Universe galaxies ($D \leq 30 $Mpc). This work builds on the previous studies of these data sets in \cite{HerreraCamus2015, Croxall2017, Abdullah2017}. 

In Section~\ref{sec:samp}, the properties of the nuclear and extranuclear star forming regions investigated and the observations used in this work are described.  Section~\ref{sec:datan} explains the data processing done to evaluate the data.  In Section~\ref{sec:res} we present the results of our analysis, comparing the measurements of \CII\ from isolated ISM phases to TIR luminosity, emission from small dust grains measured by polycyclic aromatic hydrocarbon (PAH) emission feature strength, and star formation rate.  Section~\ref{sec:con} provides the conclusions drawn from this work.

\section{Sample \& Observations} \label{sec:samp}
\subsection{KINGFISH Galaxies}
Table \ref{table:samp} provides a brief description of the properties of the galaxies included in this study.  This work uses the subset of the galaxies in the KINGFISH sample with Photoconductor Array Camera and Spectrometer (PACS) \NII~205~\micron\ and \CII~158~\micron\ spectral maps \citep{Kennicutt2011}.  The overall KINGFISH survey studied 61 Local Universe galaxies ($D<30$~Mpc) spanning a wide range of morphological types, luminosities, metallicities, and star formation activity levels.  Of these 61 galaxies, 54 were observed at the \CII~158~\micron\ line.  Within this smaller sample, \textit{Herschel} PACS far-infrared spectral cubes targeting the \NII~205~\micron\ line were acquired for galaxies with the highest far-infrared surface brightnesses.  This subsample contains 31 regions in 28 galaxies, with 24 centered on the brightest, central nuclear region of the galaxy and 7 centered on extranuclear star-forming regions in the disk of the galaxy. All of the targeted galaxies in this sample are normal star-forming galaxies in terms of infrared luminosity, with no luminous infrared galaxies (LIRGs) included, i.e., all KINGFISH galaxies have $\log_{10} (L_{\rm IR}/L_\odot) \leq 11$.   The galaxies included in this work cover a range of nebular metallicities (12+log(O/H)) spanning $\sim$8.1-- 8.7 as measured by \cite{Moustakas2010} using the calibration of \cite{Pilyugin2005}, and a distance range of 3--30~Mpc \citep{Kennicutt2011}.   The physical scale of the \CII~158~\micron\ $\sim$11\arcsec\ and the \NII~205~\micron\ $\sim$14.5\arcsec\ PSF ranges from 180--1700~pc and 200--2100~pc respectively, across the sample.  The proximity and properties of the galaxies included in this sample allows us to determine the nature of the \CII\ emission in galaxies with a wide array of spectral coverage and compare our results to the U/LIRGS studied in \cite{DiazSantos2017}.  Further, by only targeting normal, star-forming galaxies we are able to explore the behavior of the \CII\ deficit without the complicating effects of AGN or other extreme conditions.  Establishing an understanding of the processes occurring in these well-studied galaxies will lay the groundwork for understanding measurements of more extreme cases.

In addition to the KINGFISH measurements, 20 of the 28 galaxies in this sample were included in the BtP survey.  The BtP survey used the Spectral and Photometric Imaging REceiver (SPIRE) on \textit{Herschel} to obtain larger area \NII~205~\micron\ maps, extending this study to the more quiescent areas surrounding the bright star formation regions included in the KINGFISH PACS 205~\micron\ survey.  The larger BtP maps introduce 127 additional 20\arcsec\ regions within these 20 galaxies.  For more information about the choice of 20\arcsec\ regions, see Section \ref{sec:datan}.  As the BtP regions cover a wider range of conditions, we split them into those centered nearest the galaxy nuclei and those further removed from the nuclei.   We distinguish these `Inner' and `Outer' regions as those within one-quarter of $R_{25}$ and those outside of one-quarter of $R_{25}$.  The number of each type of detection are listed in Table~\ref{table:samp}.  Although the BtP regions do cover a wider range of properties, the centers of only $\sim14\%$ fall outside of one-quarter $R_{25}$.  

\begin{deluxetable*}{lccrrrrrrrc}
\tablewidth{18cm}
\tablecaption{Sample}

\tablenum{1}

\tablehead{\colhead{Galaxy} &\colhead{Alternative} & \colhead{Region} & \colhead{RA} & \colhead{Dec} & \colhead{Distance} & \colhead{$L_{\rm TIR}$} & \colhead{$T$-Type} & \colhead{BtP} & \colhead{} & \colhead{} \\ 
\colhead{} & \colhead{ Name } & \colhead{ Type} & \colhead{(J2000)} & \colhead{(J2000)} & \colhead{(Mpc)} & \colhead{$L_{\odot}$} & \colhead{} & \colhead{Data} } 
\startdata
NGC0337 &     & Nuclear & 00:59:50.200  & $-07$:34:38.00  & $19.3~$ & 1.2$\times10^{10}$ &$8$ & $\dotsb$ \\
NGC0628 & [H76]292  & Extranuclear & 01:36:45.200 & +15:47:49.00  & $7.20$ & 8.0$\times10^{9}$ & $6$ & $\dotsb$  \\
NGC1097 &  UGCA041 & Nuclear  & 02:46:19.200 & $-30$:16:28.00 & $14.2~$ & 4.5$\times10^{10}$ & $3$ & \checkmark \\
NGC1266 &   & Nuclear & 03:16:00.600 & $-02$:25:39.00 & $30.6~$ & 2.5$\times10^{10}$ & $-2$ & \checkmark \\
NGC1377  &   & Nuclear & 03:36:39.500 & $-20$:54:08.00 & $24.6~$ & 1.3$\times10^{10}$ & $-2$ & $\dotsb$ \\
IC 342      & UGC02847 & Nuclear & 03:46:48.200 & +68:05:48.00  & 3.28 & 1.4$\times10^{10}$ & $7$ & $\dotsb$ \\
NGC1482 &   & Nuclear & 03:54:38.900 & $-20$:30:09.00 & 22.6~ & 4.4$\times10^{10}$ & $-2$ & \checkmark \\
NGC2146 & UGC03429 & Nuclear & 06:18:38.100 & +78:21:23.00 & 17.2~ & 1.0$\times10^{11}$ & $2$ & $\dotsb$ \\
NGC2798 &  UGC04905 & Nuclear & 09:17:22.800 & +42:00:00.00 & 25.8~ & 3.6$\times10^{10}$ & $1$ & \checkmark \\
NGC2976 & [HK83]58 & Extranuclear & 09:47:07.300 & +67:55:56.00  & 3.55 & 9.0$\times10^{8}$ & $6$ & \checkmark \\
NGC3049 & UGC05325 & Nuclear & 09:54:49.600 & +09:16:18.00 & 19.2~ & 3.5$\times10^{9}$ &$2$ & $\dotsb$ \\
NGC3077 &  UGC05398 & Nuclear & 10:03:18.900 & +68:44:03.00  & 3.83 & 6.4$\times10^{8}$ & $10$ & \checkmark \\
NGC3351 &  M095   & Nuclear & 10:43:57.900 & +11:42:13.00  & 9.33 & 8.1$\times10^{9}$ & $3$ & \checkmark  \\
NGC3521 &  UGC06150 & Nuclear  & 11:05:48.600  & $-00$:02:05.00 & $11.20$ & 3.5$\times10^{10}$ & $4$ & \checkmark  \\
NGC3627 &  MJV 14274 & Extranuclear & 11:20:16.500 & +12:58:42.00  & $9.38$ & 2.8$\times10^{10}$ & $3$ & \checkmark  \\
NGC4254 & M099   & Nuclear & 12:18:49.500 & +14:24:55.00 & 14.4~ & 3.9$\times10^{10}$ & $6$ & \checkmark \\
NGC4321 & M100   & Nuclear & 12:22:54.900 & +15:49:19.00 & $14.3~$ & 3.5$\times10^{10}$ & $4$ & \checkmark  \\
NGC4536 & UGC07732 & Nuclear & 12:34:27.000  & +02:11:19.00 & 14.5~ & 2.1$\times10^{10}$ & 4 & \checkmark \\
NGC4569 & M090   & Nuclear & 12:36:49.800 & +13:09:47.00  & 9.86 & 5.2$\times10^{9}$ & 2 & \checkmark \\
NGC4631 & UGC07865 & Nuclear & 12:42:07.700 & +32:32:35.00  & 7.62 & 2.4$\times10^{10}$ & 8 & \checkmark \\
NGC4736 &  [HK83]004   & Extranuclear & 12:50:56.500 & +41:07:09.00  & 4.66 & 5.8$\times10^{9} $& 2 & \checkmark \\
NGC4826 & M064   &  Nuclear & 12:56:43.500 & +21:41:03.00  & 5.27 & 4.2$\times10^{9}$ & 2  & \checkmark \\
NGC5055 &  M063   & Nuclear & 13:15:49.000 & +42:01:44.00  & 7.94 & 2.2$\times10^{10}$ & 4 & \checkmark \\
NGC5457 & [HK83]033 & Extranuclear & 14:03:41.300 & +54:19:03.00  & 6.70 & 2.3$\times10^{10}$ & 7 & $\dotsb$ \\
NGC5457 & UGC09013 & Nuclear& 14:03:12.800 & +54:20:52.00  & 6.70 & 2.3$\times10^{10}$ & 7 & $\dotsb$ \\
NGC5713 & UGC09451 & Nuclear & 14:40:11.400 & $-00$:17:22.00 & 21.4~ & 3.2$\times10^{10}$ & 4 & \checkmark \\
NGC5866 & UGC09723 & Nuclear & 15:06:29.500 & +55:45:44.00 & 15.3~ & 5.7$\times10^{9}$ & $-2$  & $\dotsb$  \\
NGC6946 & UGC11597 & Nuclear & 20:34:52.300 & +60:09:13.00  & 6.80 & 8.6$\times10^{10}$ & 7 & \checkmark  \\
NGC6946 &[HK83]003 & Extranuclear & 20:35:25.400  & +60:10:00.00  & 6.80 & 8.6$\times10^{10}$ & 7 & \checkmark \\
NGC6946 & [HK83]066 & Extranuclear & 20:35:11.400  & +60:08:59.00   & 6.80 & 8.6$\times10^{10}$ & 7 & \checkmark \\
NGC7331 & UGC12113 & Nuclear & 22:37:04.000 & +34:24:53.00 & 14.5~ & 5.3$\times10^{10}$ &3 & \checkmark \\
\enddata
\label{table:samp}
\tablecomments{Sample selected based on availability of KINGFISH PACS \CII~158~\micron\ and \NII~205~\micron\ spectral maps.  Measurements of distance, $L_{\rm TIR}$, and $T$-Type are from \cite{Kennicutt2011}.}
\end{deluxetable*}

\subsection{KINGFISH-PACS Line Maps}
\begin{figure*}
\plottwo{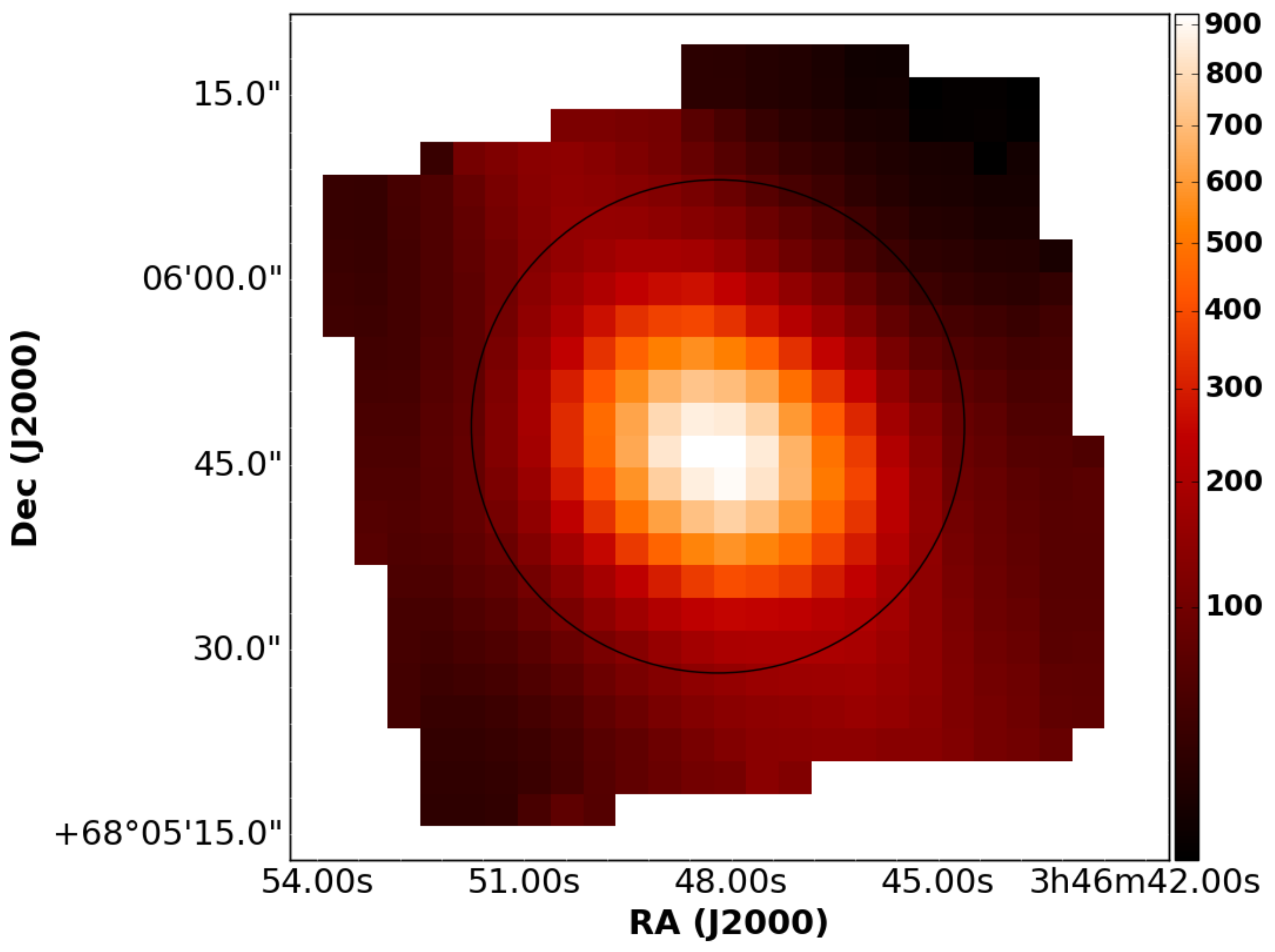}{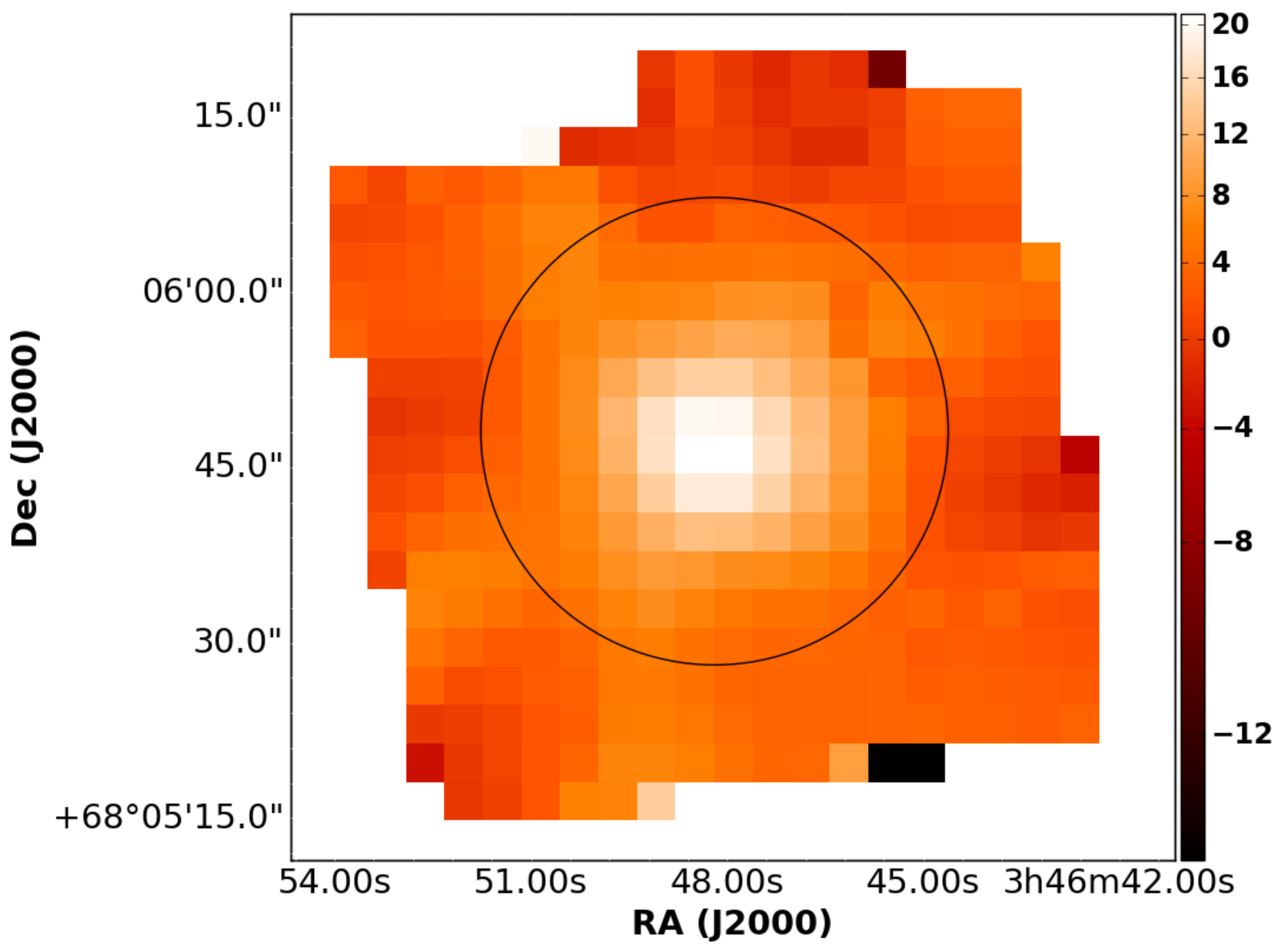}
\caption{\textit{Herschel} PACS \CII~158~\micron\ (left) and \NII~205~\micron\ (right) maps of IC~342 with the 20\arcsec\ region where fluxes were extracted marked as black circles.  Intensities are measured in nW m$^{-2}$ sr$^{-1}$.  The nuclear region of IC~342 was selected as a typical region in this study, as the measured fluxes in both \CII\ and \NII\ are reliably detected. \label{fig:ic0342}}
\end{figure*}

This work uses the KINGFISH program's \textit{Herschel}/PACS far-infrared mapped spectral observations of the 158~\micron\ and 205~\micron\ lines.  All PACS spectral maps were obtained in the Un-Chopped Mapping mode and reduced using the \textit{Herschel} Interactive Processing Environment (HIPE) version 11.2637 \citep{Smith2017}.  Standard data reductions were applied to all images and are summarized in \citet{Croxall2012}.  The resulting line maps cover a 47\arcsec\ by 47\arcsec\ square field of view with 2\farcs6 pixels in both 158 and 205~\micron and have a calibration uncertainty of 20\% and 30\%, respectively \citep{Croxall2012, Beirao2010}.  The detection of the \NII~205~\micron\ line in nearby galaxies at a spectral resolution of 150 km s$^{-1}$ is unique to the PACS instrument, making this data set invaluable for understanding the far-infrared fine structure lines in the nearby Universe \citep{Beirao2010}.  For two of the galaxies in this sample, NGC5457 and NGC6946, multiple star-forming regions were targeted; all others were observed only at the central nuclear region or at a single extranuclear star-forming region. As an example of the data from a typical nuclear pointing in this sample, the \CII~158~\micron\ and \NII~205~\micron\ maps of IC~342 can be seen in Figure~\ref{fig:ic0342}.   The flux measurements from the PACS line maps are shown in Table~\ref{table:fluxes}.

\begin{deluxetable*}{lcrrc}

\tablewidth{18cm}

\tablecaption{PACS Line Measurements}

\tablenum{2}

\tablehead{\colhead{Galaxy} & \colhead{Region} & \colhead{$L$(\CII~158~\micron) }  & \colhead{$L$(\NII~205~\micron)}  & \colhead{} \\ 
\colhead{Name} & \colhead{Type} & \colhead{(erg s$^{-1}$)} &  \colhead{(erg s$^{-1}$)}  & \colhead{} } 

\startdata
NGC0337  & Nuclear            &     774 ($\pm$ 77) $\times10^{38}$      &           <6.44$\times10^{38}$   \\ 
NGC0628  & Extranuclear    &      491 ($\pm$ 49) $\times10^{37}$      &     52.3 ($\pm$ 16) $\times10^{37}$ \\
NGC1097  & Nuclear            &      277 ($\pm$ 28) $\times10^{39}$      &      165 ($\pm$ 50) $\times10^{38}$ \\
NGC1266  & Nuclear            &      116 ($\pm$ 12) $\times10^{39}$      &        108 ($\pm$ 33) $\times10^{38}$ \\
NGC1377 &  Nuclear            &      142  ($\pm$ 15) $\times10^{38}$     &        97.0 ($\pm$ 29) $\times10^{38}$ \\
IC 342   & Nuclear               &       196 ($\pm$  20) $\times10^{38}$     &     178 ($\pm$ 53) $\times10^{37}$ \\
NGC1482  & Nuclear            &        759 ($\pm$ 76) $\times10^{39}$    &         275 ($\pm$  83) $\times10^{38}$ \\
NGC2146  & Nuclear            &       178 ($\pm$ 18) $\times10^{40}$   &         48.0 ($\pm$  14) $\times10^{39}$ \\
NGC2798  & Nuclear            &       406 ($\pm$  41) $\times10^{39}$      &       236 ($\pm$ 71) $\times10^{38}$ \\
NGC2976  & Extranuclear    &       425 ($\pm$  43) $\times10^{37}$    &         106 ($\pm$ 32) $\times10^{36}$ \\
NGC3049 &  Nuclear            &    604 ($\pm$ 60) $\times10^{38}$     &        24.4 ($\pm$  7.7) $\times10^{38}$ \\
NGC3077  & Nuclear            &     759 ($\pm$ 76) $\times10^{38}$   &          159 ($\pm$  48) $\times10^{36}$ \\
NGC3351  & Nuclear             &    443 ($\pm$  44) $\times10^{38}$  &           33.8 ($\pm$ 10) $\times10^{38}$ \\
NGC3521  & Nuclear             &     395 ($\pm$ 40) $\times10^{38}$   &          274 ($\pm$  83) $\times10^{37}$ \\
NGC3627 &  Extranuclear    &      765 ($\pm$ 77) $\times10^{38}$    &          52.5 ($\pm$ 16) $\times10^{38}$ \\
NGC4254 &  Nuclear     	   &     103 ($\pm$  10) $\times10^{39}$        &     84.5 ($\pm$  25) $\times10^{38}$ \\
NGC4321 &  Nuclear           &    107 ($\pm$  11) $\times10^{39}$      &       125 ($\pm$ 38) $\times10^{38}$ \\
NGC4536  & Nuclear           &     300 ($\pm$ 30) $\times10^{39}$     &        39.5 ($\pm$ 12) $\times10^{38}$ \\
NGC4569 &  Nuclear           &     198  ($\pm$ 20) $\times10^{38}$       &      <1.80$\times10^{38}$ \\
NGC4631 &  Nuclear          &      670 ($\pm$  67) $\times10^{38}$       &      223 ($\pm$ 67) $\times10^{37}$ \\
NGC4736 & Extranuclear   &      104 ($\pm$  10) $\times10^{38}$     &        27.7 ($\pm$ 8.4) $\times10^{37}$ \\
NGC4826 &   Nuclear         &     179 ($\pm$ 18) $\times10^{38}$     &        151 ($\pm$ 45) $\times10^{37}$ \\
NGC5055 &   Nuclear         &      227 ($\pm$ 23) $\times10^{38}$   &          227 ($\pm$ 68) $\times10^{37}$ \\
NGC5457 &  Extranuclear  &         194 ($\pm$ 19) $\times10^{38}$    &         5.29 ($\pm$ 2.9) $\times10^{37}$ \\
NGC5457 &  Nuclear          &   530 ($\pm$ 53) $\times10^{37}$        &     101 ($\pm$ 30) $\times10^{37}$ \\
NGC5713 &  Nuclear          &    349 ($\pm$  35) $\times10^{39}$      &       151 ($\pm$ 45) $\times10^{38}$ \\
NGC5866 &  Nuclear         &      224 ($\pm$ 22) $\times10^{38}$     &       43.5 ($\pm$ 13) $\times10^{38}$ \\
NGC6946 &  Nuclear          &     606 ($\pm$  61) $\times10^{38}$       &      302  ($\pm$ 91) $\times10^{37}$ \\
NGC6946 &  Extranuclear   &      216 ($\pm$   22) $\times10^{38}$     &        109 ($\pm$ 33) $\times10^{37}$ \\
NGC6946 &  Extranuclear   &       965 ($\pm$   97) $\times10^{37}$      &       157  ($\pm$ 47) $\times10^{37}$ \\
NGC7331 &  Nuclear          &    760 ($\pm$  76) $\times10^{38}$   &          91.9 ($\pm$ 28) $\times10^{38}$ \\
\enddata
\tablecomments{Luminosites are measured for 20\arcsec\ radius regions smoothed to a 20\arcsec\ FWHM PSF (see Section~\ref{sec:datan} for more information).  10\% and 30\% calibration uncertainties are included for \CII~158~\micron\ and \NII~205~\micron\ luminosities respectively.}

\label{table:fluxes}

\end{deluxetable*}

\subsection{Beyond the Peak SPIRE-FTS Line Maps}
In addition to the KINGFISH PACS \NII~205~\micron\ observations, the SPIRE-FTS on \textit{Herschel} mapped the \NII~205~\micron\ line in 20 of the KINGFISH galaxies with PACS \NII~205~\micron\ detections as part of the Beyond the Peak program.  These observations were obtained using the SPIRE-FTS intermediate mapping mode, which is a 4-point dither \citep{Croxall2017}.  These maps were calibrated using the extended-flux HIPE~v10 package.  The spectral resolution for these data is $\sim$300 km s$^{-1}$ at the 205~\micron\ wavelength. More information about the observations and data processing can be found in \cite{Pellegrini2013} and A. Crocker et al. (2019, in preparation).  The SPIRE-FTS beam size at 205~\micron\ is 16\farcs6 \citep{Makiwa2013}, and the maps produced by this survey are significantly larger than the PACS  \NII~205~\micron\ maps from the KINGFISH survey (2\arcmin~$\times$~2\arcmin\ vs 1\arcmin~$\times$~1\arcmin\ in most cases).  The inclusion of these larger area spectral maps extends the range of ISM conditions covered to cooler quiescent material surrounding the star-forming regions included in the KINGFISH sample.  Some of the important properties of these maps are listed in Table~\ref{tab:spectra}. 

\begin{deluxetable}{lllcccc}
\tablewidth{10pt}
\tablecaption{Summary of Spectral Map Data}
\tablenum{3}

\tablehead{\colhead{Telescope} & \colhead{Instrument} & \colhead{$\lambda$}  & \colhead{$\sim$PSF FWHM}  \\
 \colhead{} & \colhead{} & \colhead{\micron\ }  & \colhead{\arcsec\ }   }  
\startdata
\textit{Spitzer}  &IRS SL & 5--14  & 2--3 \\
\textit{Spitzer} & IRS LL & 14--38 & 3--10\\
\textit{Herschel}  & PACS & 158 &  11.4   \\
\textit{Herschel} &  PACS & 205 &  14.5   \\
\textit{Herschel} & SPIRE & 205  & 14.5 \\
\enddata
\label{tab:spectra}
\end{deluxetable}

\subsection{PACS Imaging}
As part of the KINGFISH survey, PACS images at 70~\micron, 100~\micron, and 160~\micron\ were obtained in scan mode for each galaxy in this sample.  These data provide valuable information about the dust temperatures of our sample and were used to determine the far-infrared colors and total infrared luminosities for our regions.  These measurements were then used to quantify the \CII\ deficit for our sample (see Section~\ref{sec:def}).  A uniform surface brightness sensitivity of $\sigma_{\rm sky} \sim$ 5, 5, and 2~MJy~sr$^{-1}$ at 70~\micron, 100~\micron, and 160~\micron\ respectively, was achieved for each region \citep{Dale2012}.  The KINGFISH PACS images have a calibration uncertainty of $\epsilon_{\rm cal, \nu} / f_{\nu} \sim 5 \%$.  For more information about these far-infrared images, see \cite{Dale2012}.  A summary of the details of these images can be found in Table \ref{tab:imaging}.

\subsection{Ancillary Infrared Observations}
Most regions in the KINGFISH sample were also included in the \textit{Spitzer} Infrared Nearby Galaxies Survey (SINGS) \citep{Kennicutt2003, Dale2006, Smith2007}.  This survey used the \textit{Spitzer Space Telescope} to obtain infrared imaging and spectroscopy for 75 nearby galaxies.  The InfraRed Spectrograph (IRS) onboard \textit{Spitzer}  obtained low-resolution ($R\sim50-100$) spectral maps in both the Short-Low (SL) (5--14~\micron) and Long-Low (LL) (14--38~\micron) modules for most of the nuclear regions in the KINGFISH sample \citep{Smith2007}.  This wavelength range includes several prominent PAH emission features.  The availability of these data allow for comparisons between the PAH emission and the \CII\ emission for the regions in our sample (see Section~\ref{sec:pah}).  The extranuclear regions targeted by the KINGFISH survey were observed by \textit{Spitzer} IRS in only the SL module.  In most cases, the larger BtP maps were covered with \textit{Spitzer}/IRS by only the LL module.  All observations were reduced with the CUBISM program to create spatially resolved spectral cubes and processed using the IRS pipeline version S14 producing and absolute flux calibration uncertainty of $\sim25\%$ \cite{Dale2006}.  For more information on these observations and data processing, see \cite{Smith2007b}.   

As part of the SINGS survey, 3.6~\micron, 8.0~\micron, and 24~\micron\ images were also obtained for each galaxy using the Infrared Array Camera (IRAC) and the Multi-band Imaging Photometer (MIPS) \citep{Rieke2004}.  These images were then processed using the MIPS Instrument Team Data Analysis Tool \citep{Gordon2005}.  Calibration errors for these images are $10\%$ at all three wavelengths \citep{Dale2005}.  Four of the galaxies in this work, IC~342, NGC~2146, NGC~3077, and NGC~5457, were not included in the SINGS sample.  For these galaxies 3.6~\micron, 8.0~\micron, and 24~\micron\ images were obtained from other archival \textit{Spitzer} surveys, namely the Local Volume Legacy \citep[LVL][]{Dale2009} and MIPS GTO \citep{Pahre2004}.  The 3.6~\micron\ and 8.0~\micron\ images provide alternative measurements of PAH feature strength for regions without both SL and LL coverage.  For the purposes of this study, the 24~\micron\ data were incorporated into the determination of the total infrared luminosities as well as in our measurements of SFR.  More information about these images can be found in Table~\ref{tab:imaging}.

\subsection{Ancillary Ultraviolet Data}
In order to determine SFR, Far-Ultraviolet maps were obtained from \textit{GALEX}. 26 of the 31 regions in this sample were imaged as part of the GALEX Nearby Galaxy Survey (NGS) \citep{GildePaz2005}, and those that were not covered by NGS were imaged as part of the GALEX All-Sky Imaging Survey (AIS) or by other programs, with the exception of NGC~1377 which has no \textit{GALEX} imaging.  Priority was given to long-exposure data when available.  Exposure times for this sample range from 110--21177~seconds, with a median exposure time of $\sim$1700~seconds.   The GALEX images have a diffraction-limited FWHM of $\sim$6\arcsec\ \citep{GildePaz2005}.  For more details on the GALEX images, see Table~\ref{tab:imaging}.

\begin{deluxetable}{llccccc}
\tablewidth{10pt}
\tablecaption{Summary of Imaging Data}

\tablenum{4}

\tablehead{\colhead{Telescope} & \colhead{Band} & \colhead{Pixel Scale} & \colhead{$\sim$PSF FWHM} \\ 
\colhead{} & \colhead{} & \colhead{\arcsec\ } & \colhead{\arcsec\ }  } 
\startdata
GALEX & FUV & 1.5 & 6.0  \\
\textit{Spitzer} & IRAC  3.6 $\mu$m &  1.2 & 1.7  \\
\textit{Spitzer} &IRAC 8.0 $\mu$m & 1.2 & 2.0\\
\textit{Spitzer} &MIPS  24 $\mu$m & 1.5 & 6.0  \\
\textit{Herschel} &PACS  70 $\mu$m& 1.40 & $5.6$\\
\textit{Herschel} &PACS  100 $\mu$m& 1.70 & $6.7$ \\
\textit{Herschel} &PACS  160 $\mu$m& 2.85 & $11.4$  \\
\enddata
\label{tab:imaging}
\end{deluxetable}

\section{Data Analysis}\label{sec:datan}
To support a consistent analysis across the wide range of wavelength information we use (.15--205~\micron), we have extracted all fluxes according to an effective 20\arcsec\ resolution.  This consistency in extraction is accomplished by taking each image (and each spectral slice within the IRS cubes) at its native resolution and, after centering on each targeted region, summing the pixel values using a Gaussian weighting $\sigma_{\rm extraction}$ determined by the difference in quadrature between the desired 20\arcsec\ and native resolutions, namely
\begin{equation}\label{magicmesh}
\sigma_{\rm extraction}^2 = \sigma_{20''}^2 - \sigma_{\rm native}^2.
\end{equation}
In this procedure the data are left at the native pixel sampling. This allows for streamlined convolution to an equivalent Gaussian resolution profile across the wide-range of wavelength images we are working with.  This method can be consistently used with both images and cubes, ensuring all data are processed in the same manner.  This process follows a similar method to that used in \cite{Contursi2002}, which found little difference between this method and other methods of smoothed flux extractions. 

\subsection{Inner and Outer BtP Measurements}
As multiple studies have found the \CII\ deficit to be a local effect \citep{Smith2017, Gullberg2018}, we use the extended SPIRE coverage to determine if there are significant differences between the behavior of the phase-separated \CII\ deficit in the inner nuclear regions of these galaxies and the more extended disks.  In order to test these differences, we divide the BtP regions into `Inner' and `Outer' regions, with Inner regions centered within $0.25R_{25}$ and Outer regions as any falling outside of this limit.  As the \NII~205~\micron\ emission is faint, only $\sim 14\%$ of the BtP regions with \NII~205~\micron\ detections fall into our definition of Outer regions.  To avoid biasing our data towards the conditions within galaxies with more regions, an average of the SPIRE detections for the Inner and Outer regions of each galaxy included in the BtP survey was performed to produce the BtP Inner and Outer measurements used throughout this study.

\subsection{Neutral Fraction of \CII\ Measurements}
To determine the fraction of the \CII\ emission originating in the neutral phase of the ISM, the following relation from \cite{Croxall2012, Oberst2006} was used:
\begin{equation}\label{eq:fneut}
 f_{\rm{\CII}, \rm{Neutral}} = \frac{ \rm{\CII\ } 158 - (R_{\rm{Ionized}} \times \rm{\NII\ } 205 ) }{\rm{ \CII\ }158 }. 
 \end{equation} 

\begin{figure}
\includegraphics[width=\linewidth]{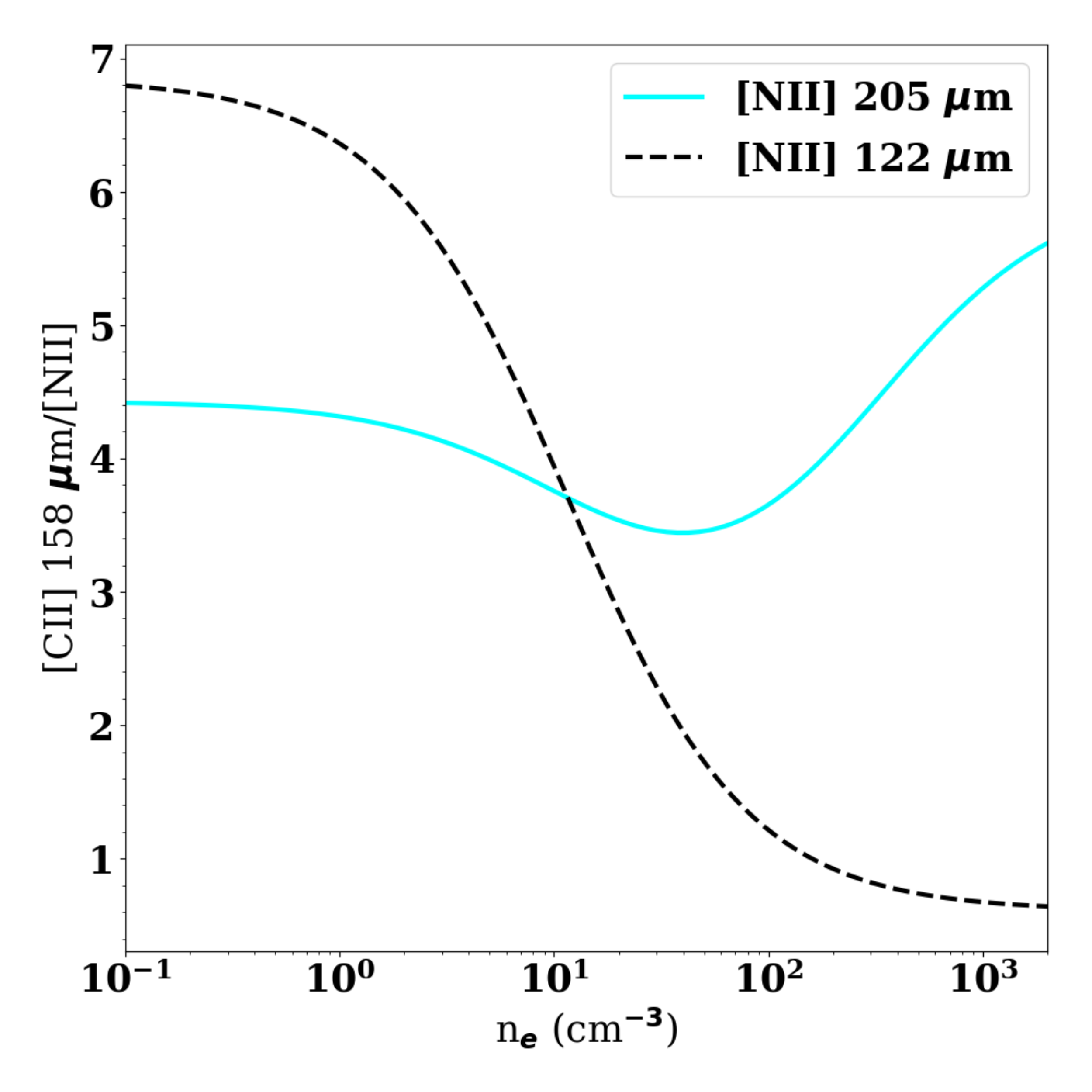}
\caption{Ratios of [CII]~158~\micron\ to [NII]~122~\micron\ and 205~\micron\ emission for ionized regions where both C$^{+}$ and N$^{+}$ are present.  Ratios determined using the method described in \cite{Croxall2017}.}
\label{fig:ciitonii}
\end{figure}

 In this equation, $R_{\rm{Ionized}}$ is the expected ratio of \CII~158~\micron\ to \NII~205~\micron\ emission in the ionized gas where both C$^+$ and N$^+$ are present, as represented by the solid cyan line in Figure \ref{fig:ciitonii}.  This ratio is derived using the collision rates of \cite{Tayal2008} for \CII\ and \cite{Tayal2011} for \NII\ and assuming Galactic gas-phase abundances of carbon and nitrogen ($1.6\times10^{-4}$ per hydrogen atom and $7.5\times10^{-5}$ per hydrogen atom respectively) \citep[see][for further information]{Croxall2017}.  It has been found that there is a slight dependence of $R_{\rm{Ionized}}$ on gas-phase abundances, but this dependance only results in shifts of $ \leq 5\%$ on measurements of $f_{\rm{\CII}, \rm{Neutral}}$ across the range of metallicities included in this sample, and therefore does not effect our results \citep{Croxall2017}.  $R_{\rm{Ionized}}$ is determined for each region individually based on electron number density measurements made using ratios of the [SIII]~18.7~\micron\ and 33.4~\micron\ lines \citep{Dale2006} or the \NII~122~\micron\ and 205~\micron\ lines \citep{HerreraCamus2016}.  Over the range of conditions covered by our sample, these two ratios both provide reliable measurements of $n_e$ \citep{Rubin1995}.  Any slight differences in calculations of $n_e$ between these two methods will not affect our results, as $R_{\rm{Ionized}}$ is nearly independent of $n_e$.  Studies of local high specific star formation rate (sSFR) galaxies have found that there is a correlation between $n_e$ and sSFR that could potentially affect the ionization parameter in the most dense regions in our study \citep{Kewley2015, Bian2016, Holden2016}, but we have not modified our $R_{\rm{Ionized}}$ calculation to account for this.

\begin{figure}
\includegraphics[width=\linewidth]{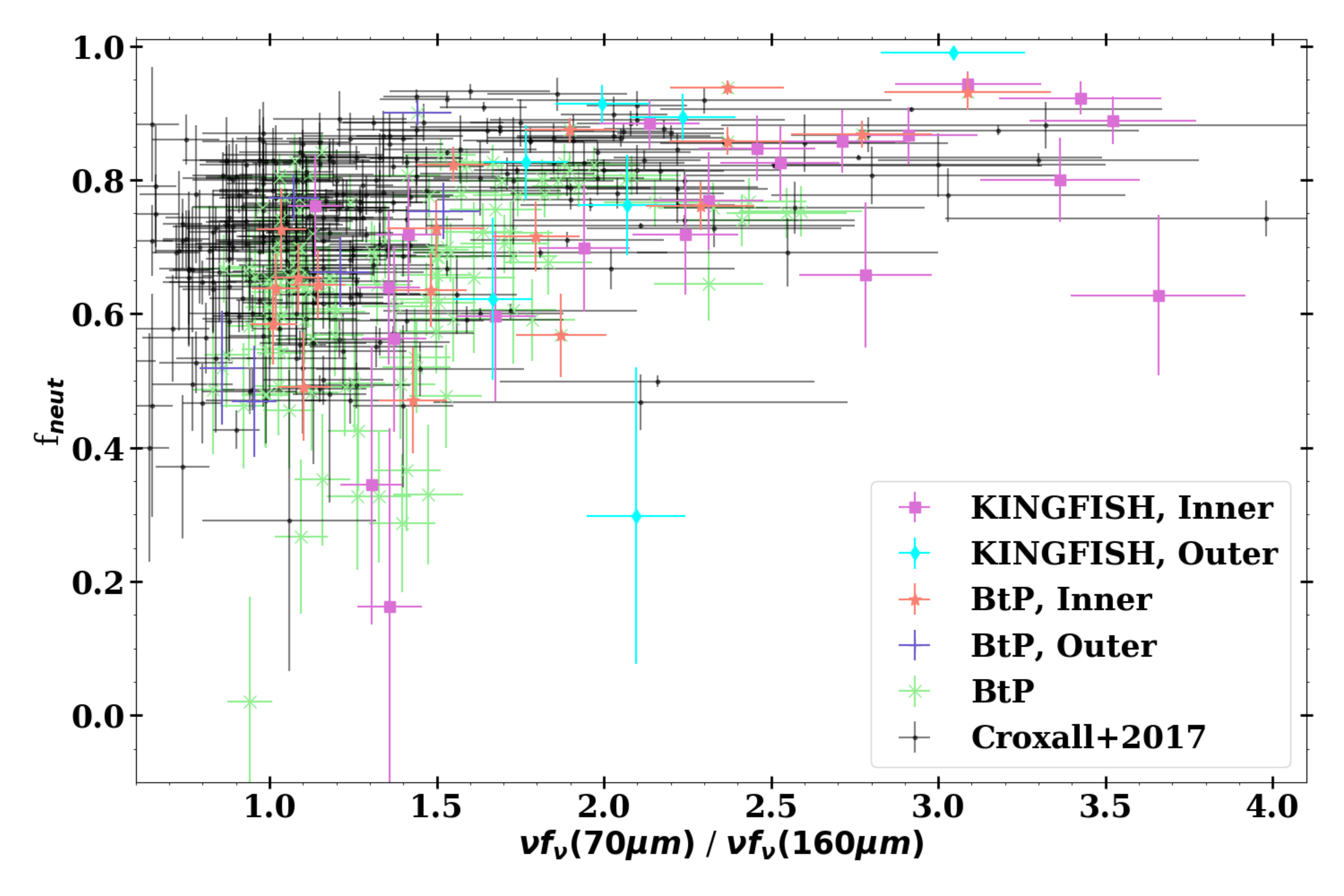}
\caption{$f_{\rm{\CII, Neutral}}$ measurements for the nuclear regions in the KINGFISH sample (magenta squares), extranuclear regions in the KINGFISH sample (cyan diamonds), averaged inner regions in the BtP sample (orange stars), averaged outer regions in the BtP sample (blue plus marks), and all individual BtP regions (green crosses), plotted along with the data from \cite{Croxall2017}, shown as grey points. Both samples show a similar pattern of increased neutral fraction with far-infrared color as measured by the $\nu f_{\nu}(70 \mu \rm{m})$/$\nu f_{\nu}(160 \mu \rm{m})$ ratio.}
\label{fig:fneut}
\end{figure}

The values of $f_{\rm{\CII}, \rm{Neutral}}$ for the KINGFISH and BtP samples are shown in Figure~\ref{fig:fneut}, plotted against $\nu f_{\nu}(70 \mu \rm{m})$/$\nu f_{\nu}(160 \mu \rm{m})$, a proxy for dust temperatures.  Also shown in Figure~\ref{fig:fneut} are the values determined using smaller subregions measured using \textit{Herschel} SPIRE data in \cite{Croxall2017}.  The $f_{\rm{\CII}, \rm{Neutral}}$ values found for this sample follow a similar trend to those in the \cite{Croxall2017} study, with an average neutral fraction of $\sim 67\%$ and a decreasing dynamic range in $f_{\rm{\CII}, \rm{Neutral}}$ for the more actively-star forming galaxies that have increased $\nu f_{\nu}(70 \mu \rm{m})$/$\nu f_{\nu}(160 \mu \rm{m})$ values.  In this and all proceeding figures, the KINGFISH regions, shown as magenta squares, cover only higher $\nu f_{\nu}(70 \mu \rm{m})$/$\nu f_{\nu}(160 \mu \rm{m})$ ratios as all KINGFISH regions are centered on warmer star-forming regions, while the wider field of view BtP data extend to the quiescent environments surrounding these star forming regions.

\vskip 2cm 

\vskip 5cm 

\subsection{PAH measurements}\label{sec:pahmeas}  
The PAHFIT program was used to determine the strength of the PAH emission features from the fluxes extracted from the IRAC spectral maps \citep{Smith2007, Gallimore2010}.  This program separates emission from PAH features, far-infrared emission lines, warm dust, and stars.  For the KINGFISH nuclear regions with IRAC spectral maps in both SL and LL large enough to cover the \CII~158~\micron\ and \NII~205~\micron\ emission, the total PAH power was determined by summing the emission from all the PAH emission features measured by PAHFIT \citep{Croxall2012}.  For regions covered only in SL (KINGFISH extranuclear regions) or only in LL (BtP regions), the total PAH power was determined using a combination of the IRAC spectral maps and the 8.0~\micron\ and 3.6~\micron\ IRAC imaging.  The bandpass used for the 8.0~\micron\ images contains one of the strongest PAH features \citep{Croxall2012}.  Stellar contributions can be removed from the 8.0~\micron\ flux through the use of the 3.6~\micron\ IRAC images.  The  total PAH power can thus be calculated photometrically using the result determined from the Local Volume Legacy (LVL) survey \citep{Marble2010}:
\begin{equation}\label{eq:pahphot}
\text{PAH}^*_{\rm Phot} = [\nu S_{\nu}(8.0) - 0.24 \times \nu S_{\nu}(3.6)].
\end{equation}
With the added information from either the SL or LL bands, this photometrically determined PAH power can be improved using the results from \cite{Croxall2012}:
\begin{equation}\label{eq:pahsl}
\text{PAH}^*_{\rm SL} = 0.497 \times [\text{PAH}^*_{\rm{Phot}} + 3.59 \times \nu S_{\nu} (11.3)] 
\end{equation}
\begin{equation}\label{eq:pahll}
\text{PAH}^*_{\rm LL} = 0.7472\times [\text{PAH}^*_{\rm{Phot}} + 3.25 \times \nu S_{\nu} (17)]  
\end{equation}
With the \textit{Spitzer} IRAC imaging and spectral maps, estimates of the total PAH emission for most regions in our sample were determined, allowing for comparisons of the PAH emission and \CII\ emission in both actively star-forming regions and the more quiescent environments surrounding these regions.  For our sample, the KINGFISH extranuclear regions total PAH emission strength is determined using Equation~\ref{eq:pahsl} and the BtP regions total PAH emission strength is determined with Equation~\ref{eq:pahll}.  These three methods produce similar measurements of total PAH emission power with a 1~$\sigma$ scatter of 0.15~dex and 0.09~dex compared to the summation of all PAH features for Equations~\ref{eq:pahll} and \ref{eq:pahsl} respectively \citep{Croxall2012}. 

\section{Results \& Discussion}\label{sec:res}

\subsection{\NII\ and \CII\ deficits} \label{sec:def}
\begin{figure}
\includegraphics[width=\linewidth]{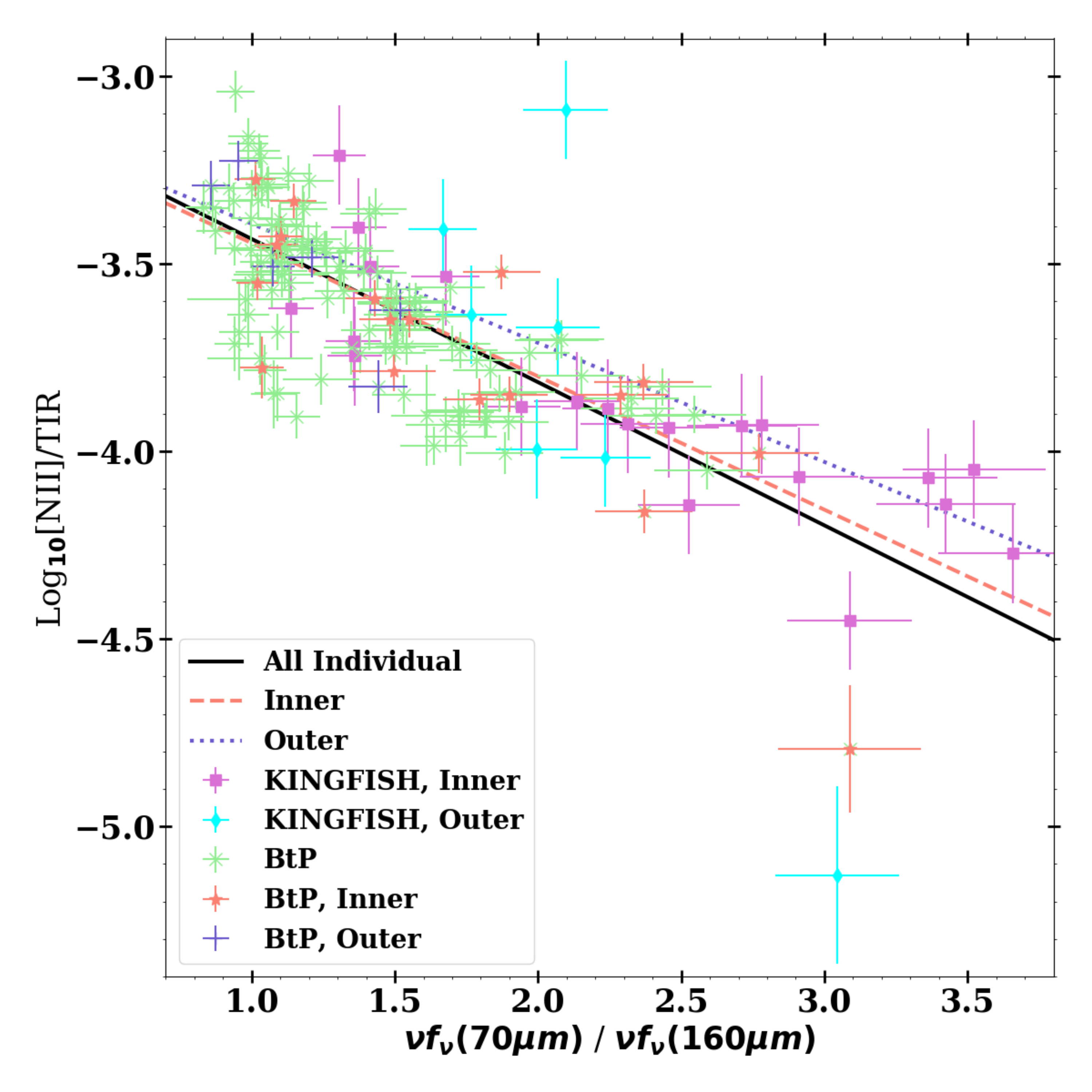}
\caption{\NII~205~\micron~/~TIR plotted against the far-infrared color measured by $\nu f_{\nu}(70\mu \rm{m}) / \nu f_{\nu}(160\mu \rm{m})$.  The decreasing trend as a function of warmer regions shows that the \NII~205~\micron\ line has a deficit in our sample.  Lines of best fit are shown for the trend found with all the individual regions in the KINGFISH and BtP samples (black line), the KINGFISH nuclear regions and the averaged BtP inner regions (orange dashed line), and the KINGFISH extranuclear regions and the averaged BtP outer regions (blue dotted line). }
\label{fig:niidef}
\end{figure}

The \NII~/~TIR measurements for our sample are displayed as a function of the far-infrared color as measured by the $\nu f_{\nu}(70\mu \rm{m}) / \nu f_{\nu}(160\mu \rm{m})$ ratio in Figure~\ref{fig:niidef}.  The $\nu f_{\nu}(70\mu \rm{m}) / \nu f_{\nu}(160\mu \rm{m})$ is used  as it is a proxy for dust temperature and increased dust temperatures often indicate increased star formation activity.  TIR luminosity was determined for each individual region using Equation 4 from \cite{Dale2014} (Equation~\ref{eq:tir} in this work) with the  \textit{Spitzer} 24~\micron\ luminosities and \textit{Herschel} PACS 70~\micron\ and 160~\micron\ luminosities:
\begin{equation}\label{eq:tir}
L_{\rm TIR} = 1.548\nu L_{\nu}(24~\mu{\rm m}) + 0.767\nu L_{\nu}(70~\mu{\rm m}) + 1.285\nu L_{\nu}(160~\mu{\rm m}).
\end{equation}
Similar to the findings of \cite{DiazSantos2017}, we find that the \NII~205~\micron\ line ratio with TIR shows a clear decreasing trend in warmer regions, and this trend holds irrespective of sampling the inner or outer portions of the galaxies.  Using the form:
\begin{equation} \label{eq:niidef}
 \log_{10}{ \frac{\rm{\NII}}{\rm{TIR}}} = m \nu f_{\nu}(70 \mu \rm{m})/\nu f_{\nu}(160 \mu \rm{m})+b
 \end{equation}
we perform a linear regression for all the individual regions in our sample, the averaged inner BtP measurements and the KINGFISH nuclear regions, and the averaged outer BtP measurements and the KINGFISH extranuclear regions.  For these and all following fits, the KINGFISH extranuclear region from NGC5457 has been ignored as it is a faint source with significant noise contamination at the edges of the image, making it an outlier in each plot (see the cyan diamond with a  $ \log_{10}{ \frac{\rm{\NII}}{\rm{TIR}}}$ just below -3.0 in Figure~\ref{fig:niidef}).  The slopes and intercepts for each fit are listed in Table~\ref{tab:niifits} and displayed in Figure~\ref{fig:niidef}.  Each fit has an equivalent slope within error, showing that the location of the region within the galaxy does not significantly effect our results.

\begin{deluxetable}{llccccc}
\tablewidth{9pt}
\tablecaption{Liner fits from Figure~\ref{fig:niidef}}

\tablenum{5}

\tablehead{ \colhead{$m$} & \colhead{$b$} & \colhead{RMS Scatter}  } 
\startdata
\multicolumn{3}{c}{All Individual Regions} \\
\hline
$-0.381[\pm 0.03] $ & $-3.05$ & 0.174  \\
\multicolumn{3}{c}{All Inner Regions} \\
\hline
 $-0.355 [\pm 0.04] $ & $-3.09$ & 0.181  \\
\multicolumn{3}{c}{All Outer Regions} \\
\hline
 $-0.317 [\pm 0.22] $ & $-3.08$ & 0.246  \\
\enddata
\label{tab:niifits}
\tablecomments{Properties of the lines of best fit for our \NII\ deficit measurements, displayed in Figure~\ref{fig:niidef}.  We divide our fits by region type, with a fit for all individual regions, a fit for the averaged BtP inner regions and the KINGFISH nuclear regions, and a fit for the averaged BtP outer regions and the KINGFISH extranuclear regions.}
\end{deluxetable}

\begin{figure*}

\includegraphics[width=\linewidth]{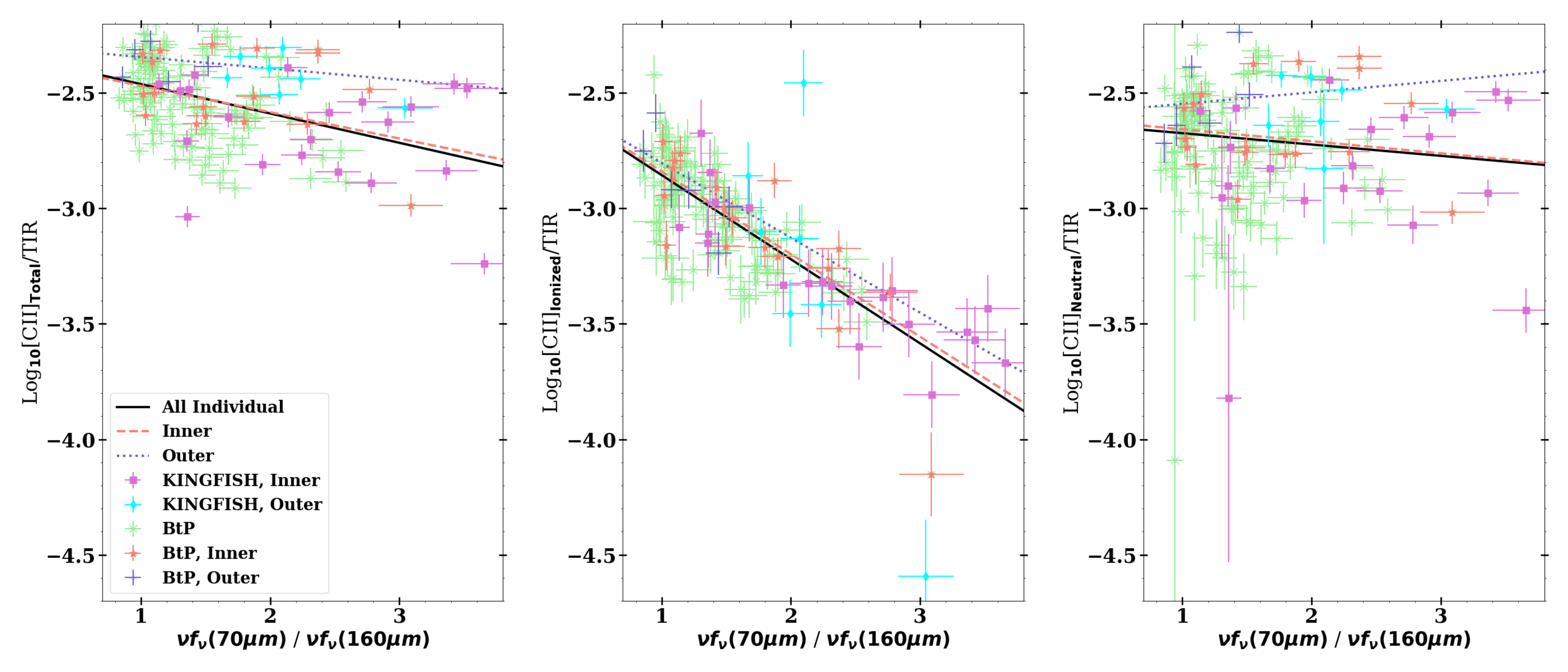}
\caption{\textit{Left}: \CII~158~\micron~/~TIR plotted against the far-infrared color measured by $\nu f_{\nu}(70\mu \rm{m}) / \nu f_{\nu}(160\mu \rm{m})$.  The \CII\ deficit is observed as a slight decrease in the \CII~158~\micron~/~TIR ratio at warmer far-infrared colors in our sample.  Our sample covers a limited range of conditions, and therefore only a small deficit effect is observed.  The lines represent predicted neutral fractions based on the relationship determined by the \CII\ emission from only the ionized phase of the ISM (shown in middle panel).  \textit{Middle:} The ratio of  \CII~158~\micron\ emission from the ionized ISM to TIR  plotted against the far-infrared color measured by $\nu f_{\nu}(70\mu \rm{m}) / \nu f_{\nu}(160\mu \rm{m})$.  \textit{Right}: The ratio of  \CII~158~\micron\ emission from the neutral ISM to TIR  plotted against the far-infrared color measured by $\nu f_{\nu}(70\mu \rm{m}) / \nu f_{\nu}(160\mu \rm{m})$.  Notice the disappearance of the observed decrease in \CII~158~\micron~/~TIR when only the neutral ISM is considered.  }
\label{fig:ciidef}
\end{figure*}

In Figure~\ref{fig:ciidef}, the \CII~/~TIR measurements are displayed as a function of $\nu f_{\nu}(70\mu \rm{m}) / \nu f_{\nu}(160\mu \rm{m})$.  In this figure, the left panel is the combined ionized and neutral \CII\ emission (\CII$_{\rm{I+N}}$), the middle panel is the \CII\ emission from only the ionized ISM (\CII$_{\rm{Ionized}}$), and the right panel is the \CII\ emission from only the neutral ISM (\CII$_{\rm{Neutral}}$).  The trend we notice in our \CII$_{\rm{Ionized}}$ measurements is a scaled version of the \NII~205~\micron\ measurements by nature of our method for determining \CII$_{\rm{Ionized}}$.   Each  \CII~/~TIR measure is fit by a linear regression of the form:
 \begin{equation} \label{eq:ciidef}
 \log_{10}{ \frac{\rm{\CII}}{\rm{TIR}}} = m \nu f_{\nu}(70 \mu \rm{m})/\nu f_{\nu}(160 \mu \rm{m}) + b.
 \end{equation}
These fits are displayed in Figure~\ref{fig:ciidef} and the properties of each fit are listed in Table~\ref{tab:ciifits}.  For our sample, we find an average of $\log_{10}{ \frac{\rm{\CII}_{\rm{I+N}}}{\rm{TIR}}} = -2.53$, $\log_{10}{ \frac{\rm{\CII}_{\rm{Ionized}}}{\rm{TIR}}} = -3.07$, and $\log_{10}{ \frac{\rm{\CII}_{\rm{Neutral}}}{\rm{TIR}}} = -2.73$.  These results match well with other studies which find the \CII\ line emission accounting for approximately $1\%$ of the total infrared emission \citep{Smith2017}.  As expected from previous work \citep{Smith2017, Croxall2012}, the combined ionized and neutral \CII\ luminosity--to--TIR luminosity ratio shows a decline with warmer far-infrared color.  As there are no extreme cases in this study, the decreasing trend in the combined ionized and neutral \CII\ / TIR ratio is slight, as shown by the slope of $-0.127$ for our line of best fit.  While the \CII\ deficit is pronounced for the \CII\ emission from the ionized ISM, it disappears when only \CII\ emission from the neutral ISM is considered, as shown in the middle and right panel of Figure~\ref{fig:ciidef}.

\pagebreak

\begin{deluxetable}{llccccc}
\tablewidth{9pt}
\tablecaption{Liner fits from Figure~\ref{fig:ciidef}}

\tablenum{6}

\tablehead{\colhead{\CII\ component} & \colhead{$m$} & \colhead{$b$} & \colhead{RMS Scatter}  } 
\startdata
\multicolumn{4}{c}{All Individual Regions} \\
\hline
Ionized+Neutral & $-0.127 [\pm 0.03] $ & $-2.33$ & 0.175  \\
Ionized &$-0.364[ \pm 0.03]$ & $-2.49$ &  0.174  \\
Neutral &$-0.049 [\pm 0.03] $& $-2.62$ & 0.272 \\
\multicolumn{4}{c}{All Inner Regions} \\
\hline
Ionized+Neutral & $-0.113 [\pm 0.05] $ & $-2.36$ & 0.197  \\
Ionized &$-0.358[ \pm 0.04]$ & $-2.48$ &  0.171  \\
Neutral &$-0.052 [\pm 0.05] $& $-2.61$ & 0.287 \\
\multicolumn{4}{c}{All Outer Regions} \\
\hline
Ionized+Neutral & $-0.050 [\pm 0.08] $ & $-2.30$ & 0.083  \\
Ionized &$-0.325[ \pm 0.26]$ & $-2.48$ &  0.339  \\
Neutral &$0.050 [\pm 0.16] $& $-2.60$ & 0.153 \\
\enddata
\label{tab:ciifits}
\tablecomments{Properties of the lines of best fit for our \CII\ deficit measurements.  A line of best fit is displayed for each component of the \CII\ emission (combined ionized and neutral, ionized, and neutral).  We further divide our fits by region type, with a fit for all individual regions, a fit for the averaged BtP inner regions and the KINGFISH nuclear regions, and a fit for the averaged BtP outer regions and the KINGFISH extranuclear regions.}
\end{deluxetable}

By separating our detections by ISM phase, we are able to narrow down the possible causes of the \CII\ deficit in our sample.  Using this method, we determine that the decreasing trend in \CII~/~TIR for warmer, more actively star-forming environments is greatly reduced when only the \CII\ emission from the neutral phases of the ISM is considered.  On the other hand, the ratio of \CII\ emission from the ionized phases of the ISM to TIR luminosity shows a steep decrease as a function of far-infrared color.  This can also be seen in the combined ionized and neutral \CII~/~TIR measurements, where the slight decrease observed is driven by the regions from the BtP survey, which have lower $f_{\rm{\CII}, \rm{Neutral}}$ values and therefore more emission from the ionized phases of the ISM (see Figure~\ref{fig:ciidef}).  This trend suggests that the cause of the \CII\ deficit occurs predominately in the ionized phases of the ISM, a conclusion that is supported by the work done in the GOALS survey \citep{DiazSantos2017}.  

This decreasing trend in the \CII\ from the ionized phases of the ISM holds for both the star--forming regions targeted in the KINGFISH study, shown as magenta squares, and in the more extended coverage from the BtP survey, shown as green crosses, and is true for both regions within $0.25R_{25}$ and those outside of this boundary, as seen in Figure~\ref{fig:ciidef} and the consistency of our measured slopes within error for each location.  The decreasing trend measured for the \CII$_{\rm I+N}$~/~TIR luminosity ratio is much shallower than the steep trend found in the \NII~205~\micron\ and \CII$_{\rm{Ionized}}$ measurements, shown in Figure~\ref{fig:niidef}.  This difference indicates that the lack of a measured deficit in the \CII\ emission from the neutral ISM is not solely caused by identical decreases in the \CII~158~\micron\ and \NII~205~\micron\ fluxes.  If this were the case we would then expect similar slopes in our \CII$_{\rm I+N}$and \NII\ deficit line fits.  Instead, there must be some differing physical processes in the neutral and ionized ISM driving the observed deficit.  

With this new insight into the nature of the \CII\ deficit, we can narrow down the possible causes of the deficit in our sample.  We suggest the \CII\ deficit is caused by changes in the fraction of UV light absorbed by dust within \HII\ regions and PDRs.  Compact~/~nuclear regions with warmer far--infrared colors have fractionally higher UV absorption by dust; $\nu f_{\nu}(70\mu \rm{m}) / \nu f_{\nu}(160\mu \rm{m})$ traces TIR/FUV for centrally concentrated distributions of dust \citep{Dale2007}, therefore higher $\nu f_{\nu}(70\mu \rm{m}) / \nu f_{\nu}(160\mu \rm{m})$ indicate regions where UV light is being proportionally more quenched by dust in PDRs and \HII\ regions.  The resulting dearth of UV emission leaking into the diffuse ISM leads to a decrease in \CII\ emission from the ionized phases of the ISM.  Previous studies have suggested that a majority of the \CII$_{\rm Ionized}$ emission must originate in the diffuse ISM as there should be little \CII\ emission originating in \HII\ regions where the \CII\ line is often thermally quenched by high temperature and densities and the availability of photons with energies above the 24.38 eV necessary to ionize C+ limit the emission of the 158~\micron\ line \citep{DiazSantos2017, HerreraCamus2018a}.  Thus, the decrease in \CII$_{\rm{Ionized}}$ as a fraction of the total infrared emission may be due to (1) a smaller fraction of the ultraviolet radiation from stars being above the Lyman limit (i.e., a deficiency in recent star formation) and (2) a smaller fraction of the H-ionizing radiation being absorbed in low-density ($n_e < 30$ cm$^{-3}$) \HII\ where \CII\ emission is not collisionally quenched, i.e. the diffuse ionized ISM.  Under these conditions, the \CII$_{\rm Neutral}$ should remain unaffected as it primarily originates in PDRs, where UV photons with energies above 11.3 eV have not been significantly quenched and are still available to ionize the carbon atoms present.  This interpretation is consistent with the explanations of the \CII\ deficit described in \cite{Abel2009} and \cite{GraciaCarpio2011} and suggests the third and fourth explanation of the deficit listed in Section~\ref{sec:intro} (i.e. the thermalization of \CII\ in \HII\ regions and increased FUV absorption leading to \OI\ becoming a major coolant in the diffuse ionized ISM) are the most likely for our sample.  It is also possible that the other mechanisms described in Section~\ref{sec:intro} have an effect on the deficit measurements for our sample, but do not explain the changes in trends between the ionized--phase \CII\ and the neutral--phase \CII\ measurements.

\subsection{\CII~158~\micron\ and PAH emission}\label{sec:pah}

PAHs are the dominant source of the photo-ejected electrons that heat the neutral ISM, which is in turn cooled through channels such as the \CII~158~\micron\ line \citep{Bakes1994, Weingartner2001}.  The mid-infrared PAH emission features are a result of vibrational and bending transitions that have been excited by the absorption of far-ultraviolet photons.  We thus compare the strength of the PAH emission features to the \CII\ emission from both the neutral and ionized phases of the ISM to better understand the microphysics underlying the gas heating by photoelectric ejection of electrons from PAHs and gas cooling by \CII~158~\micron\ emission.   Previous works have found that while the \CII~158~\micron~/~TIR decreases in warmer regions, the ratio of \CII~/PAH emission is more constant \citep{Helou2001, Croxall2012}.  By measuring this ratio in the separated ISM phases we can test the relationship between gas heating by small grains and gas cooling by \CII\ emission. 

\begin{figure*}

\includegraphics[width=\linewidth]{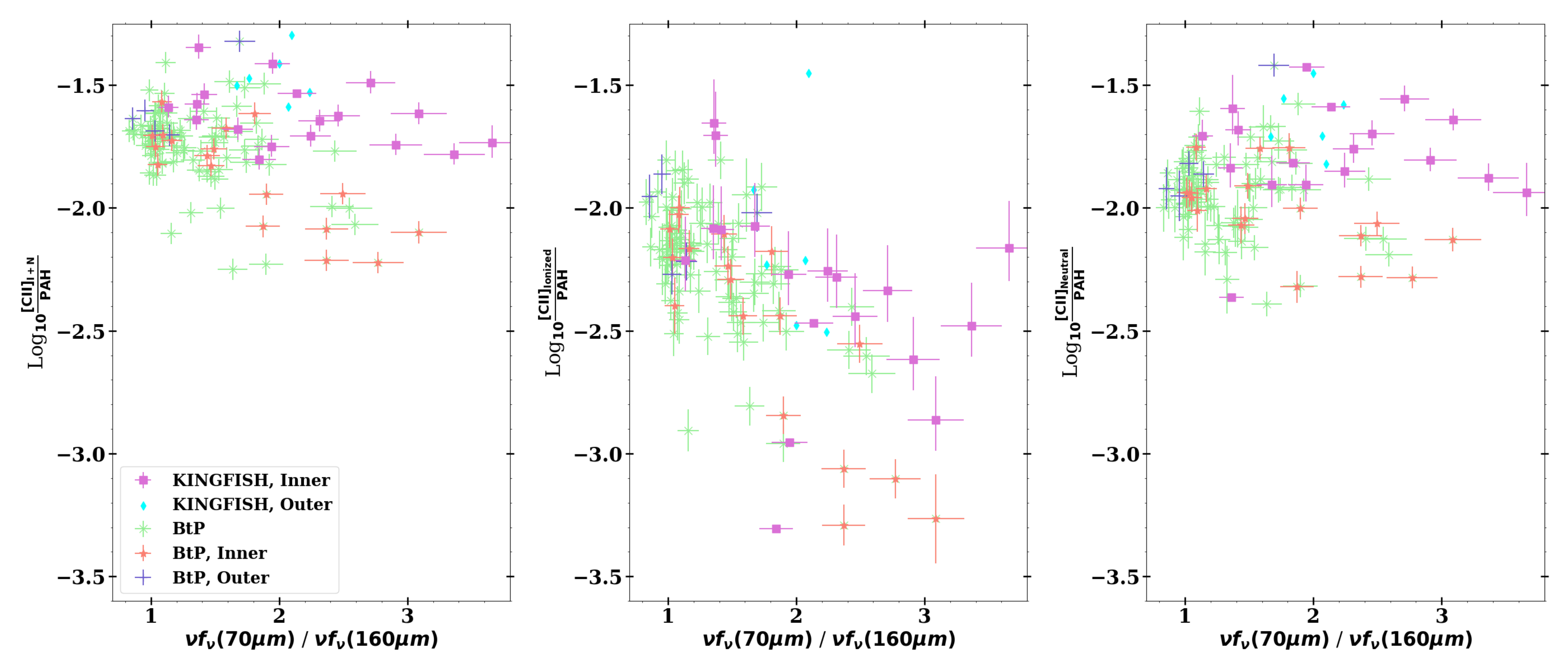}
\caption{\textit{Left:} The ratio of combined ionized and neutral \CII\ emission to PAH feature emission plotted against far-infrared color.  Notice a small decline in the warmer, more actively star-forming regions at higher  70~\micron~/~160~\micron\ ratios.  \textit{Middle:} The ratio of \CII\ emission from only the ionized ISM to PAH feature emission strength plotted against far-infrared color. \textit{Right:} The ratio of \CII\ emission from only the neutral ISM to PAH feature emission strength plotted against far-infrared color.  Notice the sharp decline in this ratio in the ionized emission and lack of decline present in the neutral emission. Markers are the same as in previous figures.}
\label{fig:ciipah}
\end{figure*}

The ratio of \CII\ luminosity to PAH feature strength for the combined ionized and neutral \CII\ emission and the \CII\ emission from only neutral and ionized phases of the ISM can be found in Figure \ref{fig:ciipah}.  KINGFISH nuclear regions where both SL and LL IRS cubes are available are shown as magenta squares, KINGFISH extranuclear regions where only SL IRS cubes are available are shown as blue squares, and BtP regions where only LL IRS cubes are available are shown as green crosses.  It should be noted that the slight separation between the different region types is likely driven by the different methods for determining PAH emission strengths, and not by any differences between the regions themselves (see Section~\ref{sec:pahmeas} for more information). 

The middle and right panels of Figure \ref{fig:ciipah} show the \CII\ to PAH emission ratio when only the \CII\ emission from the ionized and neutral ISM are considered, respectively.  Similar to results of the deficit observed when comparing the \CII\ and TIR luminosity, the ratio of \CII\ emission from only the ionized ISM to PAH emission feature strength shows a clear decrease as a function of far-infrared color, while the ratio of \CII\ emission from only the neutral ISM to PAH emission feature strength remains fairly constant across the range of far-infrared color included in this sample.  This holds for both the warmer KINGFISH nuclear and inner BtP regions as well as the slightly cooler outer BtP regions and the KINGFISH extranuclear star forming regions.

We find that the neutral \CII\ emission traces the total PAH emission well (Figure \ref{fig:ciipah}, right panel).  We interpret this result as naturally arising from the ubiquity of PAHs in the neutral ISM (e.g., PDRs) and their comparative paucity in ionized portions of the ISM such as \HII\ regions \citep[e.g.,][] {Helou2004}.  This result is consistent with \CII\ being a major cooling channel, and PAHs providing a majority of the heating through the photoelectric effect, for the neutral ISM.  Therefore, in a scenario where ISM heating is balanced by ISM cooling, \CII$_{\rm Neutral}$ emission should trace PAH emission.  As a majority of the \CII\ emission in our sample originates in the neutral ISM (Figure~\ref{fig:fneut}), the combined ionized and neutral \CII\ emission to PAH feature strength (Figure~\ref{fig:ciipah}, left panel) shows a similar trend with a slight decrease in the ratio of \CII\ to PAH emission in regions with the highest $\nu f_{\nu}(70\mu \rm{m}) / \nu f_{\nu}(160\mu \rm{m})$ values.  The ratio of \CII\ emission from the ionized phases of the ISM to PAH emission feature strength shows a decrease with respect to far-infrared color (Figure~\ref{fig:ciipah}, middle panel).  This observed decrease is due to an increasing \CII$_{\rm Neutral}$ fraction in the warmer, more actively star-forming regions (i.e., higher $\nu f_{\nu}(70\mu \rm{m}) / \nu f_{\nu}(160\mu \rm{m})$), making \CII$_{\rm Ionized}$ lower in these regions.  Such a decrease could be due to a fractionally higher absorption of UV photons within \HII\ regions for warmer, more actively star forming environments as described in Section~\ref{sec:def}.

\subsection{\CII\ as a star formation rate indicator}

\begin{figure*}
\centering
\includegraphics[width=180mm]{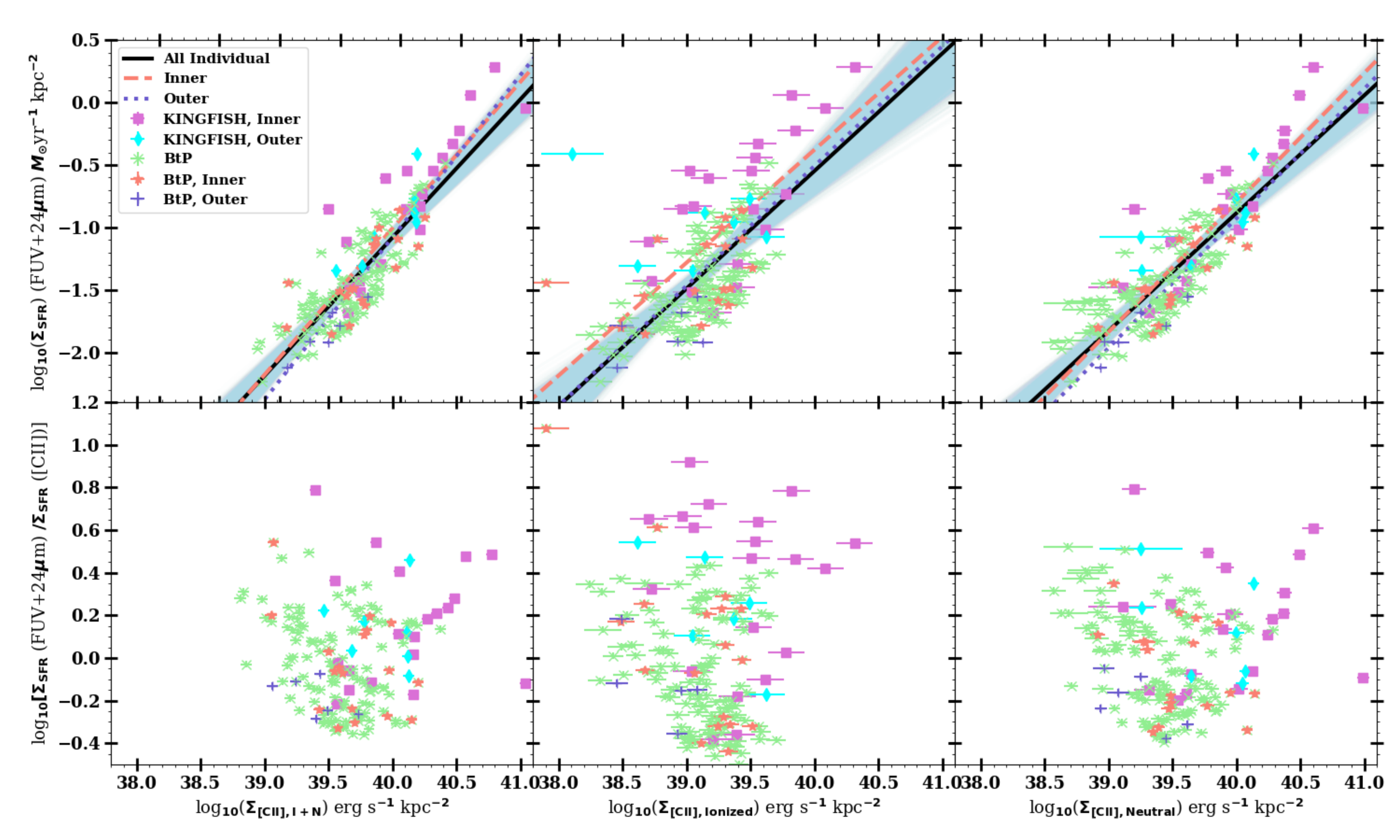}
\caption{\textit{Top}: The \CII\ surface brightness plotted vs SFR surface density for the combined ionized and neutral \CII\ emission from each region (left), only the \CII\ emission from the ionized phase of the ISM (middle), and only the \CII\ emission from the neutral phase of the ISM (right).  Black lines represent the fits determined using MCMC fitting and blue-shaded regions show the full-range of lines attempted in the MCMC fitting.  \textit{Bottom}:  The difference in the measurements of SFR using FUV+24~\micron\ measurements and using the derived combined ionized and neutral \CII\ surface brightness SFR relationship (left), \CII\ surface brightness from only the ionized ISM SFR relationship (middle), and \CII\ surface brightness from only the neutral ISM SFR relationship (right).}
\label{fig:csfr}
\end{figure*}

The top panel of Figure~\ref{fig:csfr} shows star-formation surface densities ($\Sigma_{\rm{SFR}}$) as a function of \CII\ surface brightness ($\Sigma_{\rm{\CII}}$) for the \CII\ emission from both the neutral and ionized phases of the ISM (left), the \CII\ emission arising from only the ionized phase of the ISM (middle), and \CII\ emission arising from only the neutral phases of the ISM (right).  The star formation rates were determined using the hybrid FUV+24~\micron\ local SFR indicator \citep{Hao2011, Liu2011, Calzetti2013}:
\begin{equation}\label{eq:sfr}
\text{SFR} (\rm{M}_{\odot} \rm{yr}^{-1}) = 4.6\times10^{-44} \Bigg[\frac{L(\text{FUV})}{\rm{erg \; s}^{-1}}+6.0 \frac{L(24 \text{\micron})}{\rm{erg \; s}^{-1}}\Bigg].
\end{equation}
After calculating the SFR using Equation~\ref{eq:sfr}, $\Sigma_{\rm{SFR}}$ was calculated by dividing by the de-projected area of the 20\arcsec\ regions.  The area was de-projected by dividing by $\cos{i}$  where $i$ is the inclination of the galaxy disk.  $\cos{i}$ was determined using:
\begin{equation}
cos^{2}i = \frac{(1-\epsilon)^2-q^2}{1-q^2}
\end{equation}
from \cite{Dale1997}, where $\epsilon$ is the disk's ellipticity as measured by \cite{RC3} and $q$ is an adopted intrinsic axial ratio (i.e. the ratio of the minor axis to the major axis) with a value of $q=0.13$ for galaxies of morphological class Sbc and later and $q=0.2$ for galaxies earlier than Sbc \citep{Dale1997, Murphy2018}.  The same de-projected area was used to determine $\Sigma_{\rm{\CII}}$, which is the luminosity of the \CII\ emission from each region and each separated phase divided by the de-projected area.

As found in previous studies \citep{Stacey1991, Boselli2002, DeLooze2011, DiazSantos2017, DeLooze2014, Sargsyan2012, HerreraCamus2015, Smith2017}, there are clear trends with increasing \CII\ surface brightnesses indicating increasing star formation. The lines in Figure~\ref{fig:csfr} represent the lines of best fit to our data for all individual regions as well as the combination of the KINGFISH nuclear regions and the inner BtP regions and the combination of the KINGFISH extranuclear star forming regions and the outer BtP regions, determined using the method described in \citet{Kelly2007}.  This method uses a Bayesian linear regression that takes into account both detections and upper limits.  Between 5000 and 10000 Monte Carlo Markov Chain (MCMC) steps through the parameter space are then tested to determine the best fit relationship.  The relationships found for each component of the \CII\ emission are described using Equation~\ref{eq:sfrcii} and the values displayed in Table~\ref{tab:sigmafits}.

\begin{equation}
\log_{10}{\Sigma_{\rm{SFR}} (\rm{M}_{\odot} \rm{yr}^{-1} \rm{kpc}^{-2})} =  m  \log_{10}{\Sigma_{\rm{[CII] }} }(\rm{erg} \; \rm{s}^{-1}\rm{kpc}^{-2})+ b
\label{eq:sfrcii}
\end{equation}

\begin{deluxetable}{llccccc}
\tablewidth{9pt}
\tablecaption{Liner fits from Figure~\ref{fig:csfr}}

\tablenum{7}

\tablehead{\colhead{\CII\ component} & \colhead{$m$} & \colhead{$b$} & \colhead{RMS Scatter}  } 
\startdata
\multicolumn{4}{c}{All Individual Regions} \\
\hline
Ionized+Neutral & $1.04 [\pm 0.053] $ & $-42.74$ & 0.230  \\
Ionized &$0.94[ \pm 0.085]$ & $-38.21$ &  0.333  \\
Neutral &$0.95 [\pm 0.050] $& $-38.60$ & 0.246 \\
\multicolumn{4}{c}{All Inner Regions} \\
\hline
Ionized+Neutral & $1.11 [\pm 0.112] $ & $-45.15$ & 0.270  \\
Ionized &$0.90[ \pm 0.180]$ & $-36.53$ &  0.405  \\
Neutral &$1.04 [\pm 0.105] $& $-42.29$ & 0.271 \\
\multicolumn{4}{c}{All Outer Regions} \\
\hline
Ionized+Neutral & $1.23 [\pm 0.185] $ & $-50.00$ & 0.152  \\
Ionized &$0.96[ \pm 0.400]$ & $-38.87$ &  0.314  \\
Neutral &$1.04 [\pm 0.198] $& $-42.45$ & 0.242 \\
\enddata
\label{tab:sigmafits}
\tablecomments{Properties of the lines of best fit for our $\Sigma_{\rm{SFR}}$-- $\Sigma_{\rm{\CII}}$ relationships determine using the method of \cite{Kelly2007}.  A line of best fit is displayed for each component of the \CII\ emission (combined ionized and neutral, ionized, and neutral).  We further divide our fits by region type, with a fit for all individual regions, a fit for the averaged BtP inner regions and the KINGFISH nuclear regions, and a fit for the averaged BtP outer regions and the KINGFISH extranuclear regions.}

\end{deluxetable}

To test our measurement of SFR, the \cite{Hao2011} SFR indicator determined using a lower dust attenuation coefficient, which is identical to Equation~\ref{eq:sfr} but with a proportionality constant for the 24~\micron\ luminosities of 3.89 instead of 6.0, was also applied to each region in our sample.  We find our linear fit parameters have no dependency on the coefficient we use.

The bottom panel of Figure~\ref{fig:csfr} show the differences between $\Sigma_{\rm{SFR}}$ measured with the FUV and 24~\micron\ hybrid star formation indicator (Equation~\ref{eq:sfr}) and $\Sigma_{\rm{SFR}}$ measured using the relationships we determined with all the individual regions and using the different components of the \CII\ emission (Equation~\ref{eq:sfrcii} and Table~\ref{tab:sigmafits}).  The median value of the difference for the $\Sigma_{\rm{SFR}}$ measured by the summation of the ionized and neutral \CII\ emission is $-0.024$ dex with a range of 0.80 dex to $-0.36$ dex.  The range for the differences in $\Sigma_{\rm{SFR}}$ measured by the \CII$_{\rm{Ionized}}$ surface brightness is 1.87 dex to $-0.54$ dex with a median value of $-0.058$ dex, and for the $\Sigma_{\rm{SFR}}$ measured by \CII$_{\rm{Neutral}}$ surface brightness is 0.025 dex with a range of 0.80 dex to $-0.39$ dex.  We plot these difference (logarithmic ratios) to better illustrate the scatter about our best fit lines.
 
\begin{figure*}
\centering
\includegraphics[width=180mm]{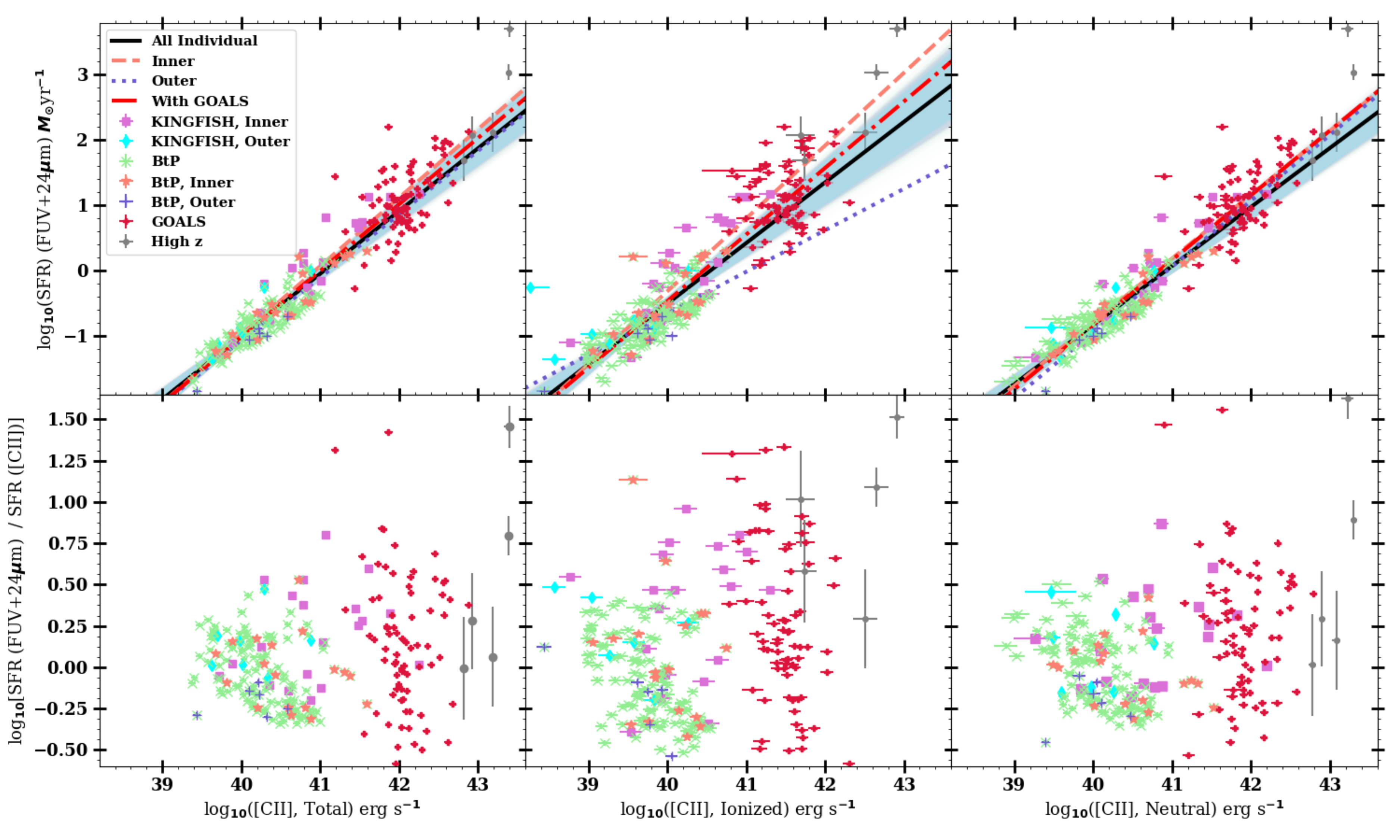}
\caption{Same as Figure~\ref{fig:csfr}, but with SFR plotted against the \CII\ luminosity from the different ISM phases.  Orange stars are measurements of LIRGS from the GOALS survey \citep{DiazSantos2017}.  Gray squares represent the limited sample of published high-redshift ($z \geq 4$) from \cite{Lu2017, Pavesi2016, Pavesi2019} where $R_{\rm{Ionized}}$ was set to four as measurements of $n_e$ are unavailable.}
\label{fig:lumcsfr}
\end{figure*}

 The \CII\ luminosity is plotted against the SFR for the combined ionized and neutral \CII\ emission and the \CII\ from the isolated ionized and neutral ISM phases in the top panels of Figure~\ref{fig:lumcsfr}.  In these plots, the LIRGS from the Great Observatory All-sky LIRG Survey (GOALS) are included to expand the range of parameter space covered \citep{DiazSantos2017}.  The LIRGS in this survey were similarly covered at the \CII~158~\micron\ line with PACS on \textit{Herschel} and at the \NII~205~\micron\ line with SPIRE-FTS on \textit{Herschel}.  More information about the observations and processing of these maps can be found in \cite{DiazSantos2013} and \cite{Zhao2013, Zhao2016a}.  The inclusion of this sample extends our study to include the more extreme infrared ($L_{\rm{IR}}~\geq~10^{11}L_{\odot}$) LIRGS that were part of the GOALS sample.  In addition to the GOALS sample, the handful of high redshift ($z \geq 4$) galaxies with \NII~205~\micron\ detections are plotted in Figure~\ref{fig:lumcsfr} \citep{Lu2017, Pavesi2016, Pavesi2019}.  As measurements of $n_e$ were unavailable for these sources, we used an average value of $R_{\rm{Ionized}}=4.0$ to determine $f_{\rm{Neutral}}$ for these sources.  None of the high-redshift sources were included in the line fitting.  Similar linear fits were preformed on these data and are shown in Figure~\ref{fig:lumcsfr}.  These trends are described using Equation~\ref{eq:sfrLcii} and the values listed in Table~\ref{tab:lumfits}.
 
\begin{equation}
\log_{10}{\rm{SFR} (\rm{M}_{\odot} \rm{yr}^{-1} )} =  m \log_{10}{L(\rm{[CII]})( \rm{erg} \; \rm{s}^{-1})} + b
\label{eq:sfrLcii}
\end{equation}

\begin{deluxetable}{llccccc}
\tablewidth{0pt}
\tablecaption{Liner fits from Figure~\ref{fig:lumcsfr}}

\tablenum{8}

\tablehead{\colhead{\CII\ component} & \colhead{$m$} & \colhead{$b$} & \colhead{RMS Scatter}  } 
\startdata
\multicolumn{4}{c}{All Individual Regions} \\
\hline
Ionized + Neutral & $0.96 [\pm 0.036] $ & $-39.46$ & 0.229  \\
Ionized &$0.93[ \pm 0.057]$ & $-37.69$ &  0.332  \\
Neutral &$0.90 [\pm 0.035] $& $-37.02$ & 0.239 \\
\multicolumn{4}{c}{All Inner Regions} \\
\hline
Ionized + Neutral & $1.05 [\pm 0.079] $ & $-42.91$ & 0.272  \\
Ionized &$1.12[ \pm 0.145]$ & $-44.98$ &  0.414  \\
Neutral &$0.98 [\pm 0.075] $& $-39.87$ & 0.270 \\
\multicolumn{4}{c}{All Outer Regions} \\
\hline
Ionized + Neutral & $0.97 [\pm 0.176] $ & $-39.86$ & 0.167  \\
Ionized &$0.64[ \pm 0.182]$ & $-26.19$ &  0.235  \\
Neutral &$1.01 [\pm 0.193] $& $-41.28$ & 0.241 \\
\multicolumn{4}{c}{With GOALS} \\
\hline
Ionized + Neutral & $1.02 [\pm 0.022] $ & $-41.64$ & 0.309  \\
Ionized &$1.02[ \pm 0.030]$ & $-41.27$ &  0.406  \\
Neutral &$0.99 [\pm 0.022] $& $-40.48$ & 0.313 \\
\hline
\enddata
\tablecomments{Properties of the lines of best fit for our SFR--L(\CII) relationships determine using the method of \cite{Kelly2007}.  A line of best fit is displayed for each component of the \CII\ emission (combined ionized and neutral, ionized, and neutral).  We further divide our fits by region type, with a fit for all individual regions, a fit for the averaged BtP inner regions and the KINGFISH nuclear regions, and a fit for the averaged BtP outer regions and the KINGFISH extranuclear regions.}

\label{tab:lumfits}
\end{deluxetable}

The bottom panels of Figure~\ref{fig:lumcsfr} show the differences between the SFR measured by the FUV+24~\micron\ hybrid star formation indicator (SFR(FUV+24), Equation~\ref{eq:sfr}) and the SFR determined using the relationships found for the \CII\ luminosity, labeled SFR(\CII) (Equation~\ref{eq:sfrLcii} and Table~\ref{tab:lumfits}).  We find a median value of $-0.036$ dex in the differences between the SFR measured using the hybrid FUV+24~\micron\ indicator and the combined ionized and neutral \CII\ luminosity, $-0.051$ dex in the differences between SFR(FUV+24) and the SFR measured by only \CII$_{\rm{Ionized}}$ luminosity, and 0.031 dex in the differences between SFR(FUV+24) and the SFR measured by only the \CII$_{\rm{Neutral}}$ luminosity.  The high redshift sources seem to follow similar trends with greater scatter, potentially due to the large uncertainties of the \NII~205~\micron\ detections.

We find that the combined ionized and neutral \CII\ emission and the \CII\ emission from only the neutral phases of the ISM trace SFR as measured by the 24~\micron\ + FUV hybrid SFR indicator with a scatter of $\sim 0.23$~dex.  The measured slope of $0.96 [\pm .036]$ for the combined ionized and neutral \CII\ luminosity--SFR relation is consistent with the relationship found by \cite{DeLooze2014}, where a slope of $1.01[\pm .02]$ was found for a sample of dwarfs, ULIRGs, AGN, and starburst galaxies and the relationship found by \cite{Pineda2014}, with a slope of $0.98[\pm .07]$ for \CII\ luminosity within the Milky Way Galaxy.  Using Equation~\ref{eq:fneut}, and our result for the SFR measured by the \CII$_{\rm{Neutral}}$ luminosity, we can write an equation for SFR measured by \CII~158~\micron\ and \NII~205~\micron\ luminosities:
\begin{equation}
 \log_{10}{\rm{SFR} (\rm{M}_{\odot} \rm{yr}^{-1})} =  0.99 \log_{10}{L ( \rm{[CII]} 158 \; \rm{erg} \; \rm{s}^{-1}) - R_{\rm{Ionized}} L(\rm{[NII]}205 \; \rm{erg} \; \rm{s}^{-1})} -40.49.
\label{eq:sfrfneut}
\end{equation}
This equation has potential to be used in both the local and high-redshift universe without a need for dust corrections.  

As the \CII\ emission from the neutral phases of the ISM accounts for most of the \CII\ emission from these regions, it is expected that the combined ionized and neutral \CII\ emission and \CII\ emission from only the neutral phases of the ISM follow similar trends, as shown by the similar slopes measured by our lines of best fit.  Although both components of \CII\ emission rise with higher star formation surface densities, the \CII\ emission from the neutral ISM shows a more tightly constrained relationship than the \CII\ emission from the ionized ISM.  This increased RMS of 0.33 for \CII$_{\rm{Ionized}}$, 0.1 dex above the RMS for the \CII$_{\rm{Neutral}}$,  is likely due to the sharp decrease in \CII$_{\rm Ionized}$~/~TIR as a function of far-infrared color.  As described in Section~\ref{sec:def}, this result could indicate that a large fraction of the \CII\ emission from ionized phases of the ISM is not coming from star-forming \HII\ regions, but instead originating in the diffuse ionized ISM \citep{DiazSantos2017, HerreraCamus2018a}.  The large scatter in the \CII$_{\rm Ionized}$--SFR relationship indicates that any attempt to use \CII\ emission as a tracer of SFR must be treated with caution in galaxies that will have a large fraction of \CII\ emission from ionized phases of the ISM, like high-redshift Ly$\alpha$ emitter galaxies.  This conclusion is supported by analysis of the kpc--resolution \CII\ detections from the SHINING survey \citep{HerreraCamus2018a}.  We also find no difference in the slopes within error when we separate our regions by location within the galaxy suggesting that these results hold in a variety of conditions. The inclusion of the GOALS sample raises the slope slightly, which we believe is due to the elevated star formation rates of these U/LIRGS which causes them to fall above the galaxy main sequence \citep{Elbaz2011, Murphy2013}.

\section{Conclusions}\label{sec:con}
With the recent availability of \NII~205~\micron\ detections in local Universe galaxies from the KINGFISH and BtP surveys, we are able to distinguish emission from the \CII~158~\micron\ line from the ionized and neutral phases of the ISM.  The sub-kiloparsec resolution of these \NII~205~\micron\ spectral maps make them an ideal resource for separating \CII\ emission by ISM phase.  Our main conclusions are:

\begin{itemize}
\item The \CII\ emission from our sample primarily originates from the neutral ISM, with an average neutral fraction of $f_{\rm{\CII}, \rm{Neutral}}$ of 67\%.  The \CII\ emission from the ionized ISM only dominates in a few regions where far-infrared color temperatures are coolest.
\item The trend of decreasing \CII~/~TIR as a function of far-infrared color, commonly referred to as the \CII\ deficit, is most prominent when only the \CII\ emission from the ionized phases of the ISM are considered, and is almost non-existent in the \CII\ emission from the neutral ISM.  
\item The differences in the behavior of the \CII\ deficit are likely due to the majority of the ionized \CII\ emission originating in the diffuse ionized ISM.  In warmer regions with increased deficit, the FUV radiation required to heat the diffuse ionized ISM is proportionally more absorbed by dust and therefore unavailable to ionize carbon, decreasing the \CII\ emission we observe from this phase.
\item The ratio of \CII\  emission from the neutral ISM to PAH emission strength is fairly constant in our sample, suggesting that in the neutral ISM gas heating is controlled by PAHs.
\item The ratio of \CII\ emission from the ionized ISM to PAH emission strength decreases sharply as a function of infrared color.  This result is consistent with a majority of the \CII\ emission from the warmer regions originating in the neutral ISM, decreasing the strength of  \CII$_{\rm Ionized}$.
\item We find the \CII\ emission from the neutral phases of the ISM traces the SFR with scatter of $\sim0.23$~dex, while the \CII\ emission from the ionized phases of the ISM trace SFR with a scatter of scatter of $\sim0.33$~dex.  The smaller scatter in the neutral \CII--SFR relationship is inherently tied to the lack of a \CII$_{\rm Neutral}$ deficit.
\item We do not find strong dependencies on spatial location within the galaxies.  However, 85\% of the regions sampled lie with $0.25R_{25}$, limiting the interpretation of this result.
\end{itemize}

The work presented here is limited to the normal star-forming galaxies observed with the PACS spectrometer in the KINGFISH survey.  Studies of the \CII\ emission in AGNs and LIRGs have found \CII--SFR relationships are more scattered in extreme conditions \citep{DeLooze2014, HerreraCamus2015}.  Despite this increased scatter for infrared-luminous and accretion-powered environments, there is reason to believe the SFR--\CII$_{\rm Neutral}$ relationship presented here will hold in a wide variety of environments.  Additional measurements of the \CII\ and \NII\ lines in LIRGs produce similar trends in the deficit behaviors for \CII\ emission from the ionized and neutral phases of the ISM \citep{DiazSantos2017}.  We plan to further investigate the samples presented, particularly with respect to any trends that may depend on quantities such as metallicity and photo-electric efficiency.  Better understanding in detail the nature of the \CII\ deficit in local galaxies, where we can disentangle the contributions from different ISM phases, is critical to interpreting \CII\ observations of galaxies at higher redshifts.

We would like to thank George Privon and Tanio D\'{i}az-Santos for enlightening discussions and contributions.  D.A.D. would like to thank IPAC/Caltech for hosting him during the beginning stages of this research.  This work was supported by NASA Headquarters under the NASA Earth and Space Science Fellowship Program, Grant \#80NSSC18K1107, as the Wyoming NASA Space Grant Consortium, NASA Grant \#NNX15AI08H.  \textit{Herschel} is an ESA space observatory with science instruments provided by European-led Principal Investigator consortia and with important participation from NASA. IRAF, the Image Reduction and Analysis Facility, has been developed by the National Optical Astronomy Observatories and the Space Telescope Science Institute.

\bibliography{js.11Oct.arxiv.CIISFR}

\end{document}